\documentclass[linenumbers]{aastex631}
\usepackage{amsmath}
\shorttitle{Circumbinary planet stability}
\shortauthors{Georgakarakos et al.}
\graphicspath{{./}{figures/}}

\begin{document}
\nolinenumbers
\title{Empirical stability criteria for 3D hierarchical triple systems I: Circumbinary planets}

\author[0000-0002-7071-5437]{Nikolaos Georgakarakos}
\affiliation{Division of Science, New York University Abu Dhabi, PO Box 129188, Abu Dhabi, UAE}
\affiliation{Center for Astrophysics and Space Science (CASS), New York University, Abu Dhabi, PO Box 129188, Abu Dhabi, UAE}

\author[0000-0002-1398-6302]{Siegfried Eggl}
\affiliation{Department of Aerospace Engineering / Department of Astronomy / NCSA CAPS, University of Illinois at Urbana-Champaign\\
Urbana, IL, USA}

\author{Mohamad Ali-Dib}
\affiliation{Center for Astrophysics and Space Science (CASS), New York University, Abu Dhabi, PO Box 129188, Abu Dhabi, UAE}

\author{Ian Dobbs-Dixon}
\affiliation{Division of Science, New York University Abu Dhabi, PO Box 129188, Abu Dhabi, UAE}
\affiliation{Center for Astrophysics and Space Science (CASS), New York University, Abu Dhabi, PO Box 129188, Abu Dhabi, UAE}



\begin{abstract}
In this work we revisit the problem of the dynamical stability of hierarchical triple systems with applications to circumbinary planetary orbits.  We derive critical semi-major axes based on simulating and analyzing the dynamical behavior of $3 \cdot 10^8$ binary star-planet configurations. For the first time, three dimensional and eccentric planetary orbits are considered. We explore systems with a variety of binary and planetary mass ratios, binary and planetary eccentricities from 0 to 0.9 and orbital mutual inclinations ranging from $0^{\circ}$ to $180^{\circ}$. Planetary masses range between the size of Mercury and the lower fusion boundary (approximately 13 Jupiter masses). The stability of each system is monitored over  $10^6$ planetary orbital periods.
We provide empirical expressions in the form of multidimensional, parameterized fits for two borders that separate dynamically stable, unstable and mixed zones. In addition, we offer a machine learning model trained on our data set as an alternative tool for predicting the stability of circumbinary planets. Both the empirical fits and the machine learning model are tested for their predictive capabilities against randomly generated circumbinary systems with very good results. The empirical formulae are also applied to the Kepler and TESS circumbinary systems, confirming that many planets orbit their host stars close to the stability limit of those systems. Finally, we present a REST API with a web based application for convenient access to our simulation data set.
\end{abstract}



\section{Introduction} \label{sec:intro}

Binary stars make up a considerable fraction of the stellar population in our galactic neighborhood \citep[e.g.][]{2010ApJS..190....1R,2012ApJ...754...44J,2022arXiv220310066O,el2024gaia}. Exoplanets, too, seem to be common in such systems as they have been discovered in both circumstellar and circumbinary orbital configurations\footnote{NASA  Exoplanet  Archive(http://exoplanetarchive.ipac.caltech.edu).}. In the former setting, a planet orbits one star of the binary, while in the latter one the planet orbits both stars. A vital part of ruling out false positives in the search for exoplanets is the assessment of whether or not predicted orbital configurations are dynamically stable \citep[e.g.][]{2021AJ....162..234K}. This is especially true for circumbinary planets since they experience significant gravitational perturbations from the stars they orbit. Stability studies can also be informative in various problems related to planet formation \citep[e.g.][]{2021MNRAS.507.3461C,2021AJ....161..211K} or habitability \citep[e.g.][]{2021FrASS...8...44G,2022MNRAS.511.4396G}.

The problem of determining stable orbit configurations with three gravitating bodies is one of the classical problems in Celestial Mechanics. Over the centuries, many have attempted to find solutions to this problem \citep[for a review see e.g.][]{2008CeMDA.100..151G}. While general analytical solutions for the gravitational three body problem do exist, they can be impractical \citep[e.g.][]{sund}. Hence, authors have continued investigating the stability of configurations of particular astronomical and astrophysical interest, such as triple stellar and planetary systems with the aim of deriving suitable criteria for determining the dynamical fate of those systems. A variety of methods and tools have been used to achieve this goal, i.e. analytical and semi-analytical methods, numerical approaches and more recently Machine Learning \citep{1999ASIC..522..385M,2001MNRAS.321..398M,2015ApJ...799....8S, 2017MNRAS.466..276G, 2018MNRAS.476.5692L, 2020AJ....159...80Q, 2022MNRAS.516.4146V}.

Numerical solutions of ordinary differential equations have been the backbone
of investigating the stability of circumbinary planets for several decades.
Almost half a century ago, \cite{1977AJ.....82..753H} carried out a limited number of numerical simulations of three body systems consisting of a stellar binary and a planet. Based on his numerical results, he derived an empirical condition for identifying stable planetary orbits in stellar binaries.  
A few years later, \cite{1986A&A...167..379D} investigated the problem of stable and unstable circumbinary motion in the the context of the planar elliptic restricted three body problem
with equal mass primaries. He integrated numerically a number of circumbinary orbits for 500 binary periods and a planetary orbit was considered stable if its eccentricity remained smaller than ${0.3}$ throughout the whole integration time. He derived two empirical formulae representing the Lower Critical Orbit (LCO), below which all planetary orbits were unstable and the Upper Critical Orbit (UCO), above which all planetary orbits were stable.  This work was extended to binaries with unequal stellar components a few years later \citep{1989A&A...226..335D} and to non-coplanar orbits by \cite{2003A&A...400.1085P}. Finally, the work of \cite{1986A&A...167..379D} was recently extended to retrograde orbits \citep{2021NewA...8401516H}. 



The most widely used empirical criterion has been the one presented in \cite{1999AJ....117..621H}. Therein, the authors investigated the dynamical stability of planets in binary systems, either in circumstellar or in circumbinary configuration.  They performed numerical simulations of massless particles on initially circular and prograde orbits around the binary or around one of the stars, in the binary plane of motion and with different initial orbital longitudes. A variety of binary masses and eccentricities was considered.  For the circumbinary case, the binary mass ratio was taken in the range 0.1-0.9 and the binary eccentricity in the range 0-0.7.  The simulation time was set to 10000 binary periods. If a particle survived the whole integration time at all initial longitudes, then the system was classified as stable.  The semi-major axis closest to the binary at which the massless particle was stable at all initial orbital longitudes was called the critical semi-major axis and was given by
\begin{equation}
a_h=[1.60+5.10e_b-2.22e_b^2+4.12M_b-4.27e_bM_b-5.09M_b^2+4.61e_b^2M_b^2]a_b,
\label{holmaneq}
\end{equation}
where $M_b=m_2/(m_1+m_2)$ ($m_1$ and $m_2$ being the masses of the two stars), $a_b$ is the semi-major axis of the stellar binary and $e_b$ is its eccentricity.  In many cases, however,  'islands' of instability were noticed at values greater than the critical semi-major axis.
This was mainly due to the choice the authors of that study made in terms of how they defined the critical semi-major axis. 

The work of \cite{1999AJ....117..621H} was subsequently extended by \cite{2018ApJ...856..150Q}.  In that work, a number of circumbinary systems was numerically simulated with many parameter ranges widened, i.e the binary mass ratio ranged between 0.01 and 0.5 with a step of 0.01, the binary eccentricity was extended to values between 0 and 0.8 with a step of 0.01 and the integration time was set to $10^5$ binary periods. Finally, the planetary mean anomaly was sampled between $2^{\circ}$ and 180$^{\circ}$ with a step of $2^{\circ}$. The definition of stability remained the same as in  \cite{1999AJ....117..621H}. \cite{2018ApJ...856..150Q} provided two fits for the critical semi-major axis of similar form to the one in \cite{1999AJ....117..621H}.



In order to better understand the dynamics of coorbital planets around stellar binaries, \cite{2023A&A...680A..29A} performed numerical simulations of planets moving in the plane of the binary. They considered the same binary mass ratios as \cite{1999AJ....117..621H}, but extended previous studies by simulating planetary orbits with eccentricities up to 0.9. However,  \cite{2023A&A...680A..29A} only sampled binary eccentricity values up to 0.5. Moreover,
the planet was always initialized at the same position with respect to the binary star. Therefore, the potential effect of different initial mean or true anomalies could not be assessed. The integration time was $10^5$ binary orbits and a system was considered unstable if the planetary semi-major axis exhibited a variation of more than $20\%$ compared to its initial value. \cite{2023A&A...680A..29A} constructed the following best fit stability criterion: 
\begin{equation}
a_{ad}=[1.36+5.79e_b-5.87e_b^2+1.99M_b-3.14M_b^2+(1.85-2.10e_b^2+3.0e_bM_b)e_p]\frac{a_b}{1-e_p}.
\label{adeleq}
\end{equation}

\cite{1999ASIC..522..385M,2001MNRAS.321..398M} approached the problem of stability of hierarchical triple systems
by drawing an analogy between
chaotic energy exchange in the binary-tides problem and stability against escape in the three body problem.  
They derived the following formula for the critical value of the outer pericentre distance ${R_{p}^{crit}}$:
\begin{equation}
\label{marda}
R_{p}^{crit}=2.8a_b\left[(1+\frac{m_3}{m_1+m_2})\frac{1+e_{out}}{(1-e_{out})^
{\frac{1}{2}}}\right]^{\frac{2}{5}}(1-0.3\frac{I_m}{180})
\end{equation}
where $m_3$ is the mass of the more distant body of the system, ${e_{out}}$ is its orbital eccentricity and $I_m$ is the mutual inclination of the two orbits given in degrees. 
If ${R_{p}^{crit}\leq R_{p}^{out}}$, then the system is considered to
be stable. The term in parentheses that includes the mutual inclination is a heuristic correction applied 
to the original formula to account for the increased stability \citep{2004RMxAC..21..156A}.
The factor of $2.8$ was also determined empirically.
As stated in \cite{2001MNRAS.321..398M} the criterion ignores  a weak dependence on the inner eccentricity  and inner mass ratio. Numerical tests have shown that equation (\ref{marda}) works well for a wide range of parameters, but it was not tested for systems with planetary masses, perhaps, because the authors were mainly interested in using the formula in star cluster simulations \citep{2004RMxAC..21..156A}.

The study at hand is motivated by the fact that all the above circumbinary stability criteria only cover parts of the parameter space. Also, the varying definitions of stability used in previous works, if not interpreted carefully, can result in misclassification of circumbinary planetary orbits as stable while they are actually unstable and vice versa.  
In this work we aim to extend and homogenize the results of previous studies on the dynamical stability of circumbinary planetary orbits. We remedy the limitations and inconsistencies that arise from combining stability estimates from different works by carrying out a self-consistent set of numerical simulations over longer timescales, namely $10^6$ orbital periods of the
planet instead of the binary. Similarly to \cite{1986A&A...167..379D}, we construct empirical fits for the upper critical border, above which all starting positions for the planet along its orbit remain stable over the integration time and the lower critical border, below which the planetary orbit is unconditionally unstable. The most notable differences of this work compared to the research cited above are: i) we perform a comprehensive scan of 3D angular momentum directions of the planet. We investigate coplanar and non-coplanar (even perpendicular) configurations with prograde and retrograde orbits;
ii) we investigate systems where the planetary body has mass, i.e. it is not approximated as a massless test particle, iii) we conduct a comprehensive scan of orbital eccentricities, where the planet can be initially on a circular or on an elliptic orbit (with up to 0.9 eccentricity),  
iv) we study the dynamics of systems for substantially longer, namely for $10^6$ planetary orbital periods. That is at least 100 times longer than \cite{1999AJ....117..621H}. And finally, 
v) we make our stability catalog available to the community in several convenient ways, including through a web-portal and an online application programming interface (API).

The rest of this article is structured as follows: in section 2 we explain the method and setup of our numerical experiments. In section 3 we present our results along with the empirical stability formulae derived by our simulation outcome data. In section 4 we compare our fits against results from random simulations. In section 5, we use Machine Learning to study the problem of circumbinary stability, while in section 6 we apply our fits to real exoplanetary systems discovered by Kepler and TESS. Section 7 briefly describes the online tools we have developed that make our results widely accessible. Finally, we conclude this article with a discussion and a summary of our results.


\section{Methodology} \label{meth}

\begin{figure}
\begin{center}
\includegraphics[width=95mm,height=80mm]{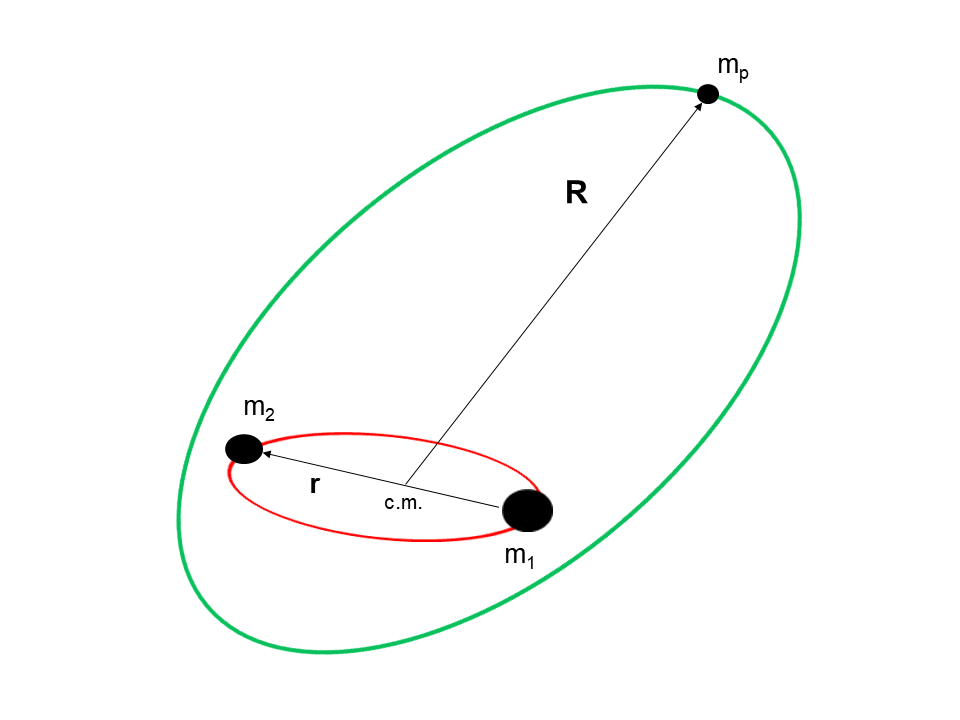}
\caption{A schematic representation of a circumbinary orbit.}
\label{fig1}
\end{center}
\end{figure}

A hierarchical triple system consists of three bodies interacting through gravity. The motion of such a system can be pictured as the motion of an ``inner binary", and an ``outer
binary".  The latter consists of the third body that orbits the centre of mass of the other two bodies (the inner binary). The problem under investigation here, i.e. the dynamical stability of circumbinary planetary orbits is a special case of a hierarchical triple configuration.  The inner binary consists of two stars, while a planet orbits their center of mass. Figure \ref{fig1} is a schematic representation of such a system. Here, $r$ is the distance between the two stars with masses $m_1$ and $m_2$, respectively, and $R$ is the distance of the planet ($m_p$) to the center of mass of the two stars.
In order to study the problem of dynamical stability of circumbinary planets, we made extensive use of numerical simulations. At the heart of our simulations is the regularized, symplectic integrator with time transformation developed in \cite{1997CeMDA..67..145M}. This code is not only symplectic in nature, which guarantees adequate conservation of system energy and angular momentum, it is also one of the few symplectic algorithms that can handle highly eccentric orbits and strong, localized gravitational interactions without loss of accuracy.  
The code uses Jacobi coordinates, i.e. it calculates the relative
position and velocity vectors of the stars and the planet at every time step.
The systems are normalized such that the gravitational constant $G=1$ and the masses of the binary stars $m_1+m_2=1$. The initial semi-major axis of the binary was also normalized to unity in corresponding units.
We convert the output of the integrator to a set of Keplerian orbital elements for the binary stars and the planet. 

All systems consist of a stellar binary and a planet on a wider orbit around the 
binary center of mass. There were no restrictions in the initial orbital configurations, i.e. we investigated circular and eccentric orbits, coplanar and non-coplanar orbits and none of the three bodies was taken to be massless.
All the bodies were treated as point masses and Newtonian gravity was the only effect considered. 

\subsection{Sampled parameter space}
Our aim is to be as comprehensive as possible in this study, which entails sweeping all relevant parameters.
We make use of two mass parameters, 
\begin{equation}
M_b=\frac{m_2}{m_1+m_2} \hspace{0.1cm}\mbox{and}\hspace{0.1cm} M_p=\frac{m_p}{m_1+m_2}.
\end{equation}
The mass parameter space was sampled as follows
\begin{equation}
M_b \in \{0.5, 0.3, 0.1, 0.05, 0.02, 0.01\}
\end{equation}
and
\begin{equation}
M_p \in \{10^{-2}, 10^{-3}, 10^{-4}, 10^{-5}, 10^{-6}, 10^{-7}\}.
\end{equation}
As our reference plane, we consider the initial orbital plane of the binary with the x-axis pointing at the direction of the binary longitude of pericenter. Hence, the binary pericenter started at zero in all cases. The choice of the initial mutual inclination between the orbital planes of the binary and that of the planet covers both prograde and retrograde orbits. We sampled values between $0^{\circ}$ and $180^{\circ}$ with a step of $18^{\circ}$, i.e. 
\begin{equation}
I_m \in \{0^{\circ}, 18^{\circ}, 36^{\circ},54^{\circ}, 72^{\circ}, 90^{\circ},108^{\circ}, 126^{\circ}, 144^{\circ}, 162^{\circ}, 180^{\circ}\}.
\end{equation}
For non-coplanar orbits, the longitude of the ascending node of the planet $\Omega_p \in \{
0^{\circ},\, 90^{\circ}\,\text{and}\,180^{\circ}\}$.  The planetary longitude of pericenter $\varpi_p$ (when dealing with coplanar orbits) and argument of pericenter $\omega_p$ (when working with three dimensional orbits) were given the same values as $\Omega_p$.
We considered initially circular and eccentric orbits for both the binary and the planetary orbit. Thus, the eccentricities were sampled as
\begin{equation}
e_b, e_p \in \{0, 0.1, 0.2, 0.3,0.4,0.5,0.6,0.7, 0.8, 0.9\},
\end{equation}
where $e_p$ is the planetary eccentricity.
Initially, the planet was placed at eight different positions with 
\begin{equation}
f_p\in \{0^{\circ}, 45^{\circ}, 90^{\circ},135^{\circ}, 180^{\circ}, 225^{\circ},270^{\circ}, 315^{\circ}\},
\end{equation}
where $f_p$ denotes the true anomaly of the planet. When the planet was on an intially circular orbit, it was started at the same angular positions around the stellar binary. 
For eccentric binaries, we used $f_b=0^{\circ}$ and $f_b=180^{\circ}$, $f_b$ being the true anomaly of the stellar binary.   

For a given set of initial parameters, we started the planet with a semi-major axis $a_p$ as close as possible to the binary (with the only requirement being that, initially, the planetary pericenter distance had to be greater than the binary apocenter distance).  Then the system was simulated progressively for higher values of $a_p$ with a resolution of 0.1.  The procedure was stopped when we could determine the stability status of the system for that set of initial conditions. Finally, the integration time was set to $10^6$ orbital periods of the planet.

\subsection{Working definition of dynamical stability}
\label{stability}
In the context of this work, a system was considered to be unstable if at least one of the following conditions was satisfied:
a) either the binary or the planetary orbital eccentricity exceeded unity, b) orbit crossing occurred, c) $a_b/a_{b_0} \leq 0.001$ or $a_b/a_{b_0} \geq 100$, d) $a_p/a_{b_0} \geq 1000$,
%
%
where $a_{b_0}$ is the initial binary semi-major axis. A system was classified as dynamically stable when the numerical simulations showed no sign of instability for any initial position of the planet on its orbit over the full time interval, i.e. $10^6$ orbital periods of the planet.  
{\bf Of course, our definition of stability is one among many, as we saw for example in the previous paragraphs when we talked about some earlier results, or as it can be seen elsewhere \citep[e.g.][]{1984CeMec..34...49S}}.  

For each set of parameters ($M_b, M_p, I_m, e_b, e_p, \Omega_p, \omega_p, \varpi_p$) we recorded three different stability regimes-areas i) stable motion for all initial true anomaly/angular position combinations, ii) mixed stable-unstable motion with at least one true anomaly/angular position combination
being stable and at least one being unstable and iii) unstable motion for all initial true anomaly/angular position combinations.  For a given set of parameters, the semi-major axis of the planet where the transition between two of the above mentioned areas of stability-instability behavior takes place is called the critical semi-major axis  $a^{cr}$. Hence, we define two such critical semi-major axes: the outer (or upper) critical semi-major axis, which is
the border between regimes (i) and (ii) and the inner (or lower)  critical semi-major axis, which is the border between regimes (ii) and (iii). This stability classification scheme was introduced in \cite{1986A&A...167..379D}, and is motivated by the fact that nonlinear dynamical systems display extreme sensitivity to changes in initial conditions in regions that are in - or close to chaotic - domains.


\section{Results} \label{sec:resu}

\begin{figure}
\begin{center}
\includegraphics[width=88mm]{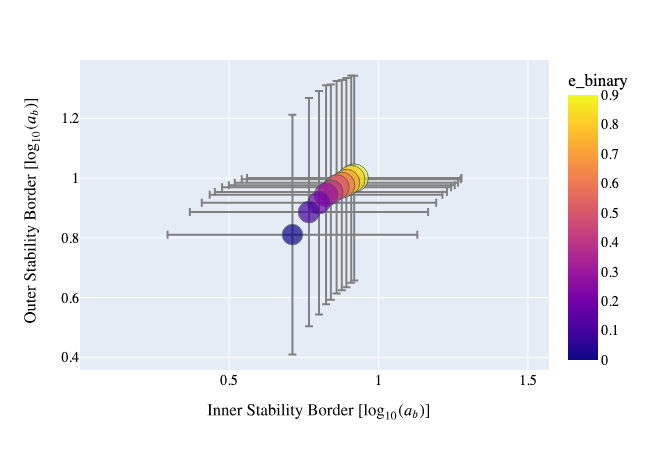}
\includegraphics[width=88mm]{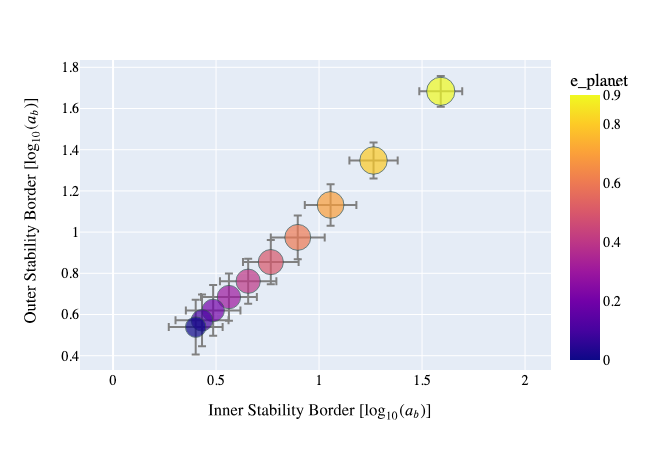}\\
\includegraphics[width=88mm]{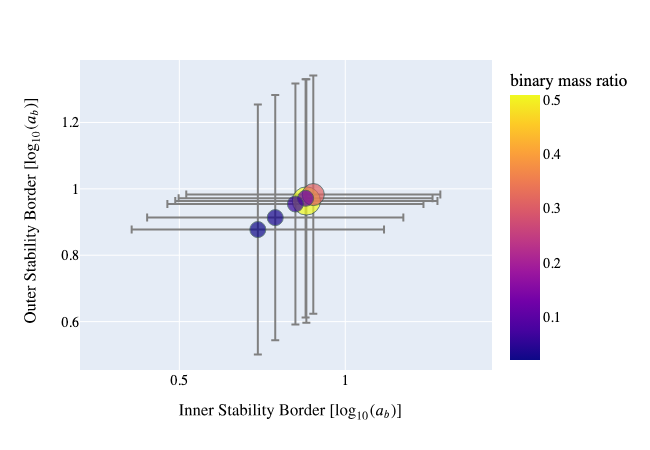}
\includegraphics[width=88mm]{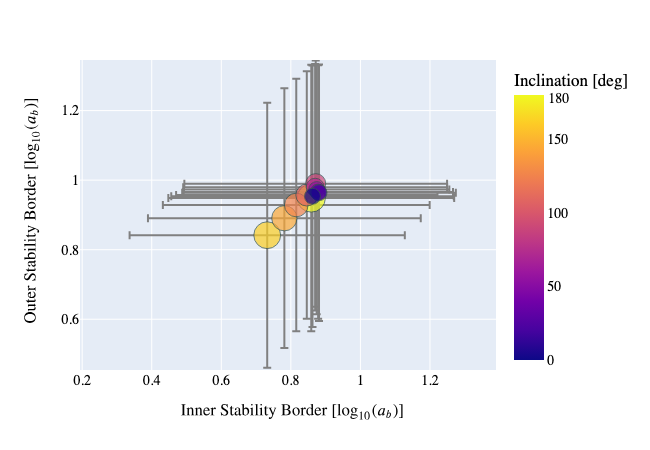}
\caption{Mean and standard deviation of outer vs inner stability borders in units of $\log_{10}$ of the binary semi-major axis. The color scale refers to the binary orbit eccentricity (top left), the planet's orbital eccentricity (top right), the binary mass ratio (bottom left), and the mutual inclination (bottom right).
Stability limits depend strongly on the planetary orbital eccentricity which accounts for most of the variance in the system. Stability borders also show roughly the same sensitivity to the binary star orbital eccentricity, the binary mass ratio as well as the inclination of the system.}
\label{fig2}
\end{center}
\end{figure}

\begin{figure}
\begin{center}
\includegraphics[width=88mm]{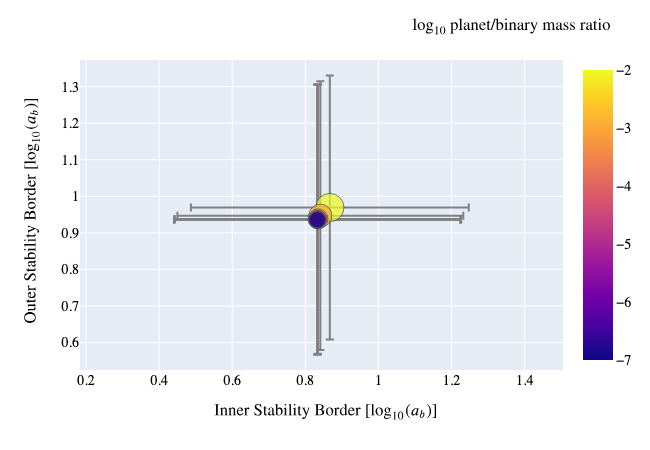}
\includegraphics[width=88mm]{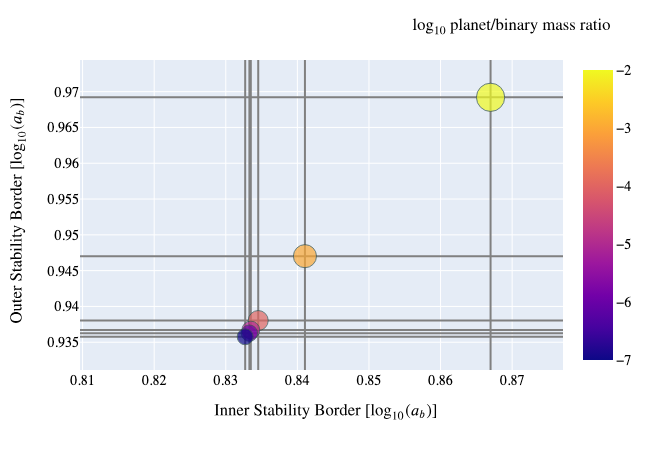}\\
\includegraphics[width=88mm]{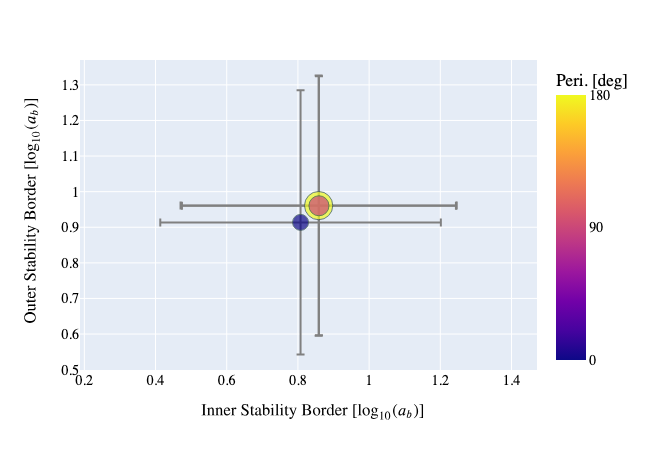}
\includegraphics[width=88mm]{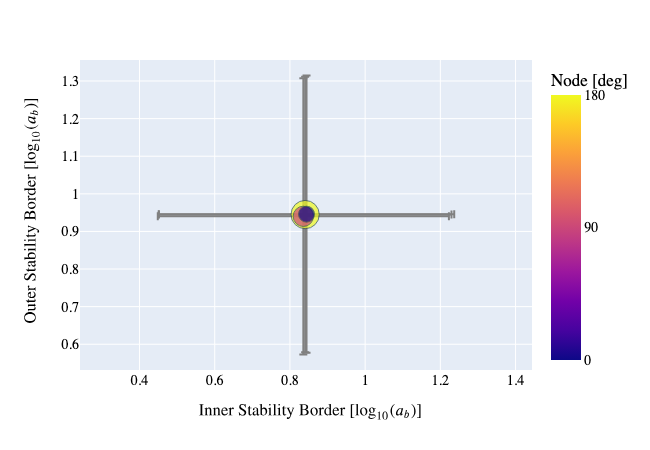}
\caption{Same as Figure \ref{fig2}, but for the planet to binary mass ratio $M_p=m_p/(m_1+m_2)$ (top left), a zoomed-in plot of the same (top right) the pericenter (bottom left), and the longitude of the ascending node (bottom right). In the parameter regime we have chosen for this study, the planet's mass does not substantially affect the stability limits. Aligned pericenters lead to lower instability in a system. The relative position of the nodes does not significantly impact the location of stability limits.}
\label{fig3}
\end{center}
\end{figure}

Our numerical study of 307,330,101 system configurations resulted in a database of critical semi-major axes values for the planet in terms of the parameters and initial conditions described in the previous section. Although circumbinary stability borders are by no means random, it can be illustrative to investigate the distribution of critical semi-major axes. A graphical representation of our results can be found in Figures \ref{fig2} and \ref{fig3} which contain plots of the mean values and standard deviations of the distribution of the outer critical semi-major axis against the inner critical semi-major axis, marginalized over all parameters except the ones displayed on the color axis.

\begin{figure}
\begin{center}
\includegraphics[width=88mm]{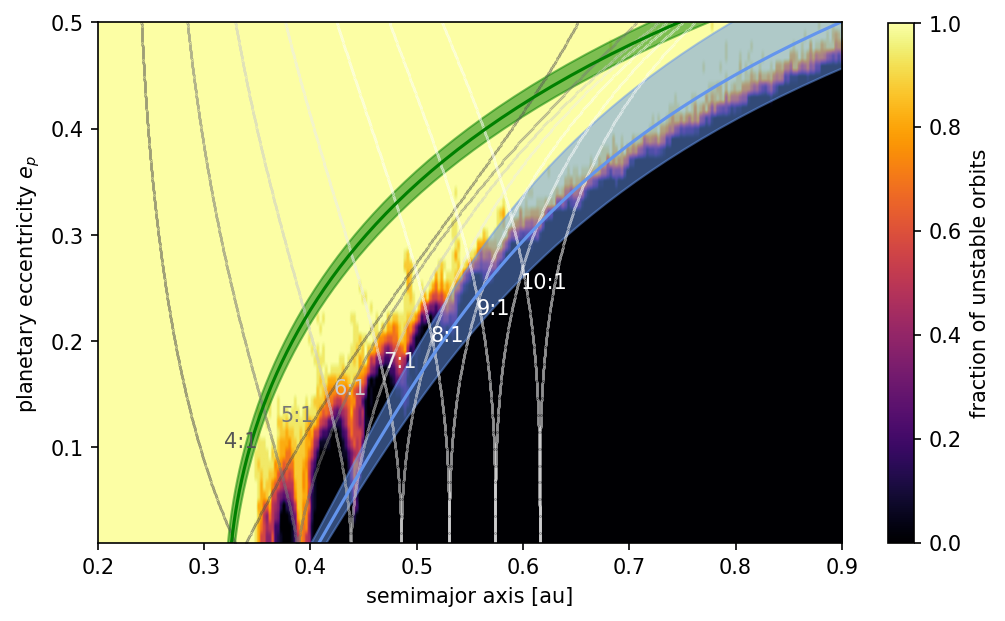}
\caption{Stability map of the TOI-1338c planet.  The parameters of the system were taken from \cite{2023NatAs...7..702S}. TOI-1338c is assumed to be the only planet in the system in a coplanar configuration. The integration time was 10000 planetary periods. Mean motion resonances between the binary and the planet are shown in gray. The green and blue lines with 3$\sigma$ uncertainty regions correspond to the empirical fits presented in equation (\ref{fits}).}
\label{fig4}
\end{center}
\end{figure}

The parameter that had the greatest effect on the stability of the systems was the planetary eccentricity (top right panel of Figure \ref{fig2}).  Higher planetary eccentricity values (especially $e_p>0.7$) yield higher values for the two critical borders. That effect was stronger when was combined with moderate to high values of the binary eccentricity. This is to be expected  since one of the main drivers of dynamical instability in the hierarchical three body problem is resonance overlap \citep{chirikov1979}.  The resonance width in such configurations is proportional to both eccentricities and has a strong dependence on the outer eccentricity when the latter becomes large \citep[e.g.][]{2008LNP...760...59M,mardling2013}.  An example of that phenomenon can be seen in Figure \ref{fig4}, where we construct a stability map for the TOI-1338 system.

The binary mass ratio and the mutual inclination seem to have a moderate effect on the location of the two stability borders. As seen in the bottom left panel of Figure \ref{fig2}, the smallest value for the binary mass ratio, $M_b=0.01$ appears to lead to larger areas of stable circumbinary motion. As we progress to higher values of the binary mass ratio, the stable area starts to shrink and it reaches its minimum when the inner binary consists of comparable mass bodies. A hint for this kind of behavior is given by secular evolution. According to \citet{2015ApJ...802...94G} we know, for instance, that the maximum eccentricity for a circumbinary orbit  $e_p^{max}$ is proportional to  $m_1m_2(m_1-m_2)/(m_1+m_2)^3$  or,
if we express that in terms of the binary mass ratio, $e_p^{max}\propto(2M_b^2-3M_b+1)M_b$.
That means smaller values of $M_b$ yield smaller values of maximum $e_p$. Lower planetary eccentricities, in turn, lead to more stable systems.

\begin{figure}
\begin{center}
\includegraphics[width=85mm,height=60mm]{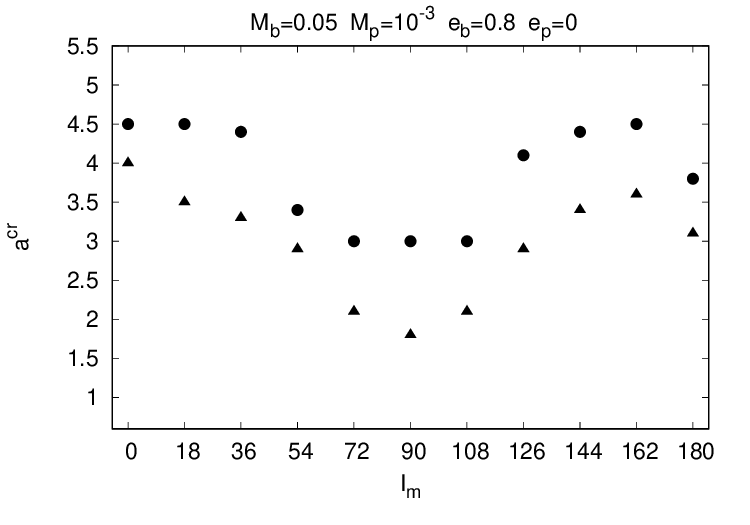}
\caption{Critical semi-major axis against mutual inclination for a system with $M_b=0.05, M_p=10^{-3}, e_b=0.8$ and $e_p=0$. The triangles represent the inner critical border while the circles indicate the outer one.}
\label{fig5}
\end{center}
\end{figure}

When considering the effect of the mutual inclination on circumbinary stability, our simulation outcome confirmed that, generally, retrograde orbits near inclinations of $160^{\circ}$ appear to be more stable, i.e. the stability borders were closer to the stellar binary compared to a prograde system with the same parameters \citep[e.g.][and bottom right panel of figure \ref{fig2}]{2008CeMDA.100..151G,2013NewA...23...41G}. However, and this was seen in previous studies as well \citep[e.g.][]{2011MNRAS.418.2656D,2020MNRAS.494.4645C}, there were cases where the behavior of the stability border as a function of the mutual inclination did not follow that trend. Figure \ref{fig5} provides such an example.  The stability border behavior seen in figure \ref{fig5} may be linked to a nodal libration of the planetary orbit around $\pm90^{\circ}$ induced by the stellar binary.  This dynamical behavior has been noted and discussed in several studies before \citep[e.g.][]{2009MNRAS.394.1721V,farago2010high,2017AJ....154...18N}.

While all bodies in our simulations were massive, 
we focused on relatively low mass external perturbers that do not exceed one percent of the mass of the inner binary. Consequently, the mass of the outer companion did not have a significant effect on the stability borders (top row of Figure \ref{fig3}). Unsurprisingly, the highest value of $M_p=0.01$ showed the larger impact, especially for systems
where the binary members had a significant mass difference.  

Finally, the slowly evolving angles, i.e. the longitude of the ascending node and the argument of pericenter of the planetary orbit did not substantially affect the location of our critical borders as seen in the bottom row of Figure \ref{fig3}.  Of course, there were cases for which there was a significant deviation between the smallest and largest stability border values for different combinations of the slow varying angles. This difference in the dynamical evolution of a planetary systems depending on the orientation of the orbits, however, is something well known in Celestial Mechanics \citep[e.g.][]{2004Icar..168..237M,hadjidemetriou2006symmetric}.   

\begin{figure}
\begin{center}
\includegraphics[width=120mm]{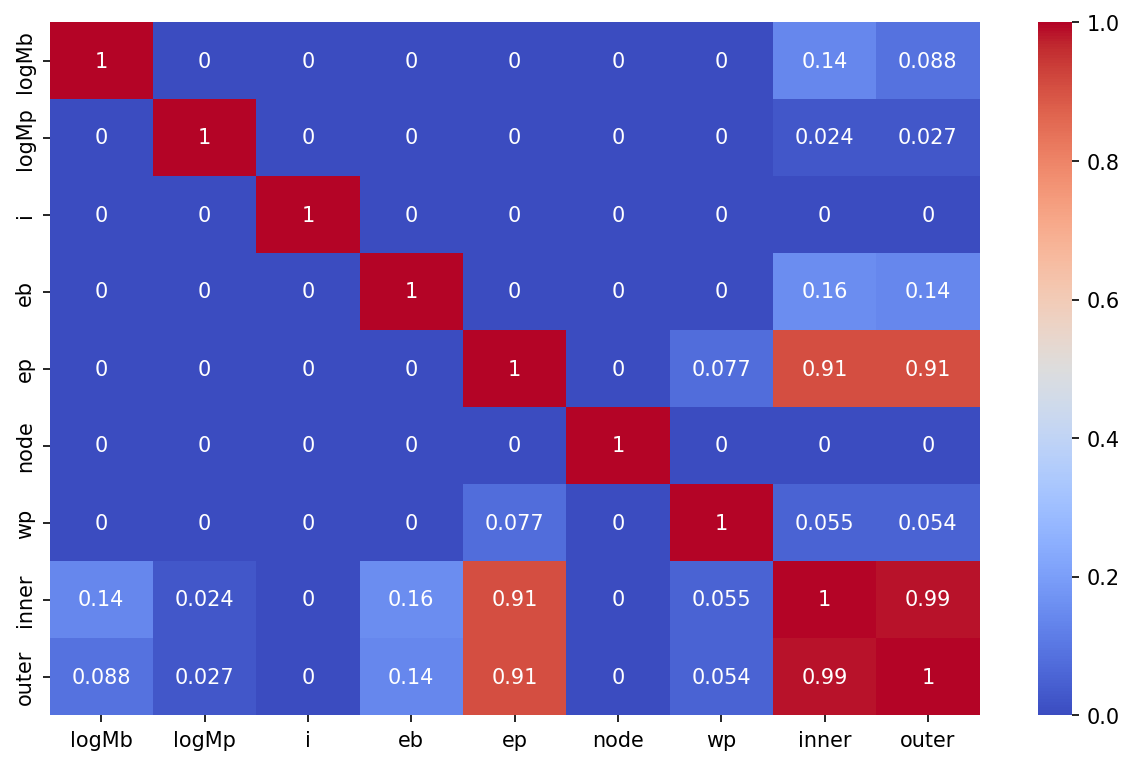}
\caption{Correlation matrix for the simulation dataset. The color indicates Pearson's correlation coefficient. 
 The logarithms of the masses are base 10 and i denotes the mutual inclination. See text for more details.}
\label{fig6}
\end{center}
\end{figure}

Constructing a correlation matrix is another approach to quantifying interdependencies between the variables in our simulation dataset. The correlation matrix contains Pearson's correlation coefficients for all model variables, i.e. the covariance of pairs of two variables divided by the product of their standard deviations. 
A correlation matrix for our dataset is presented in Figure \ref{fig6}. The inner and outer border are almost perfectly correlated, as is evident from the diagonal structure in the mean values in Figures \ref{fig2} and \ref{fig3}. The correlation matrix also indicates that the influence of the planetary eccentricity $e_p$ dominates the inner and outer stability limits.

\begin{figure}
\begin{center}
\includegraphics[width=59mm,height=50mm]{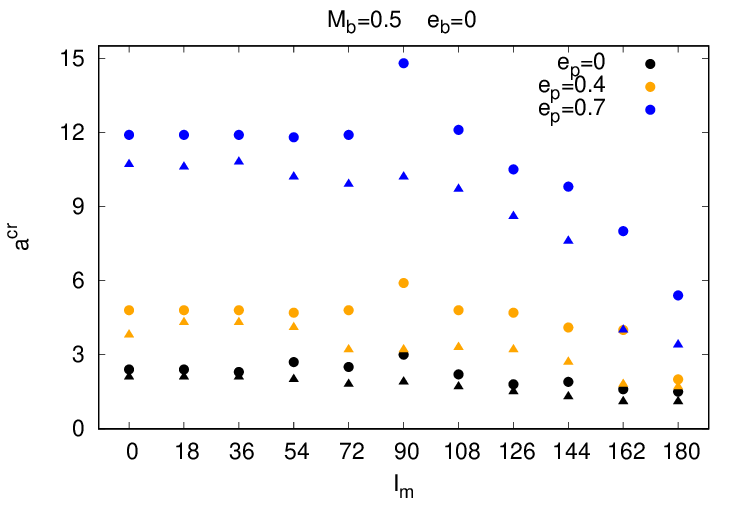}
\includegraphics[width=59mm,height=50mm]{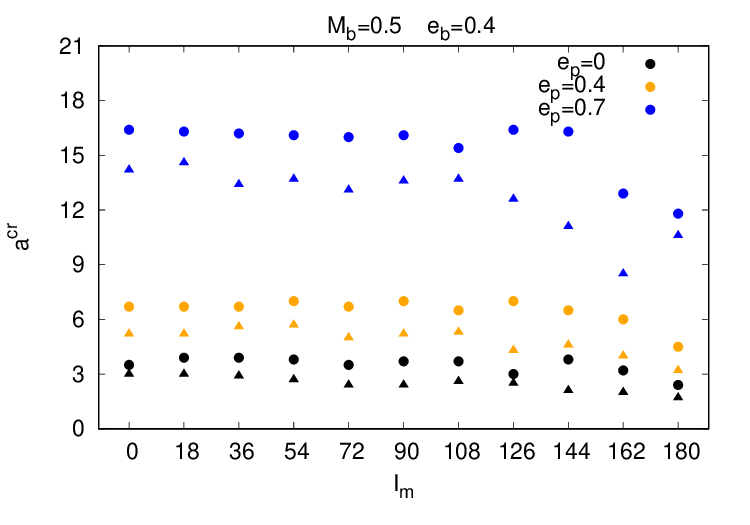}
\includegraphics[width=59mm,height=50mm]{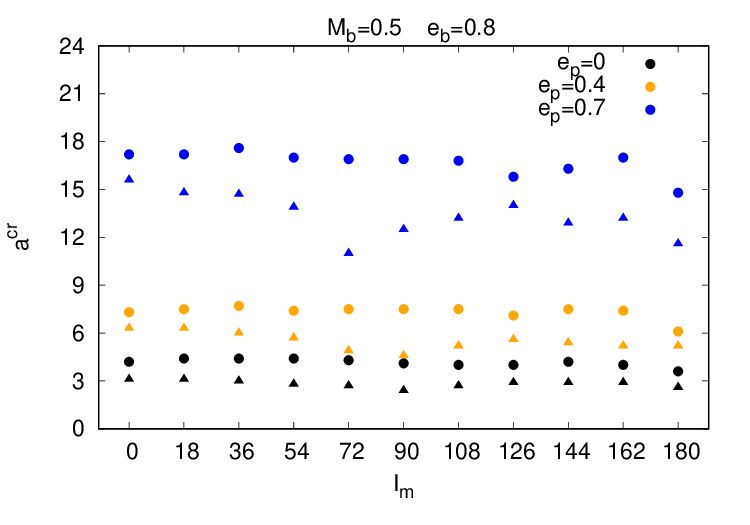}
\includegraphics[width=59mm,height=50mm]{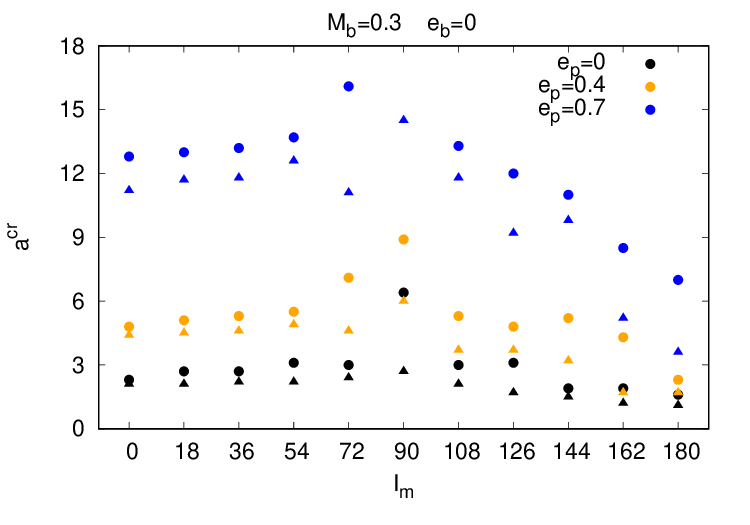}
\includegraphics[width=59mm,height=50mm]{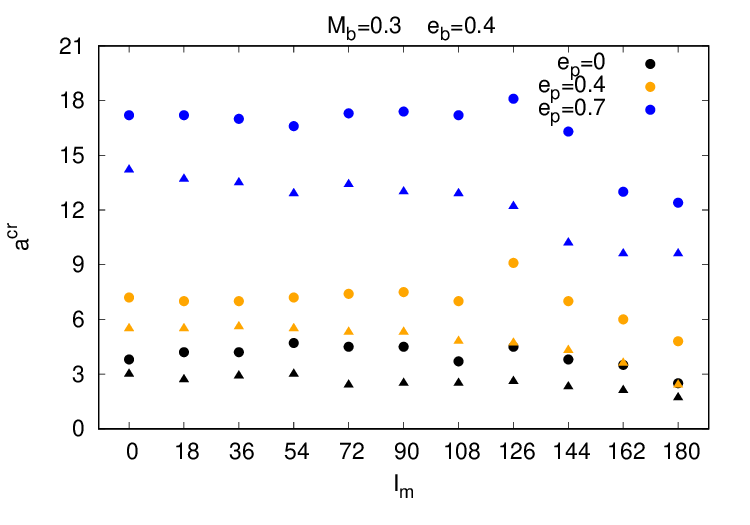}
\includegraphics[width=59mm,height=50mm]{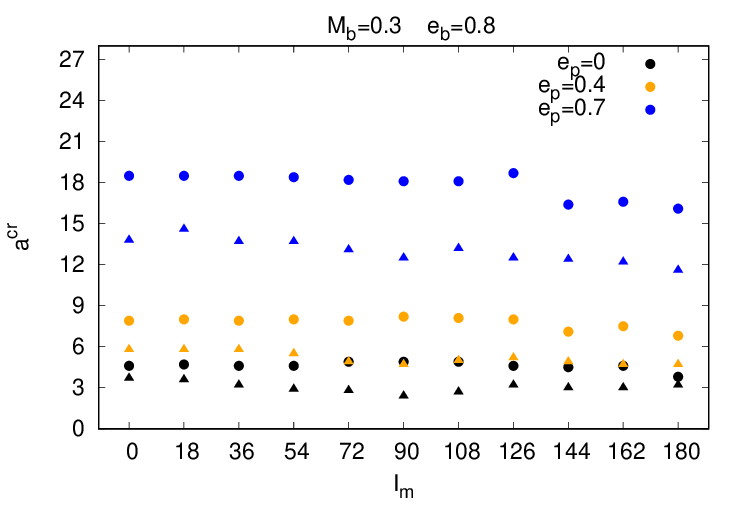}
\includegraphics[width=59mm,height=50mm]{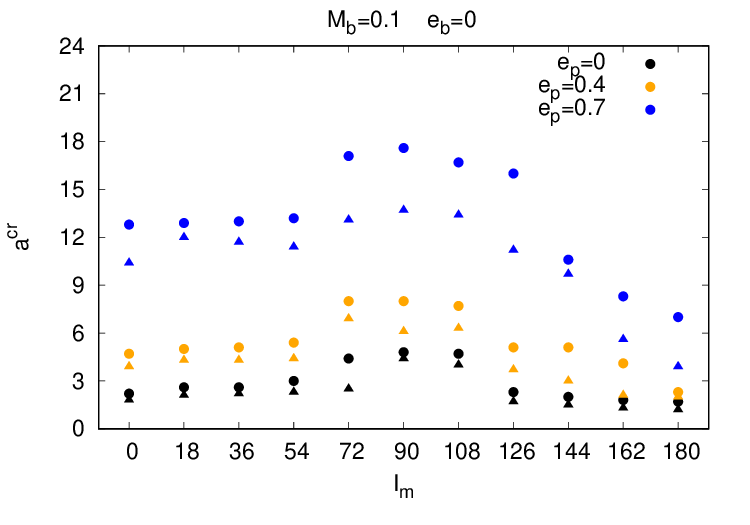}
\includegraphics[width=59mm,height=50mm]{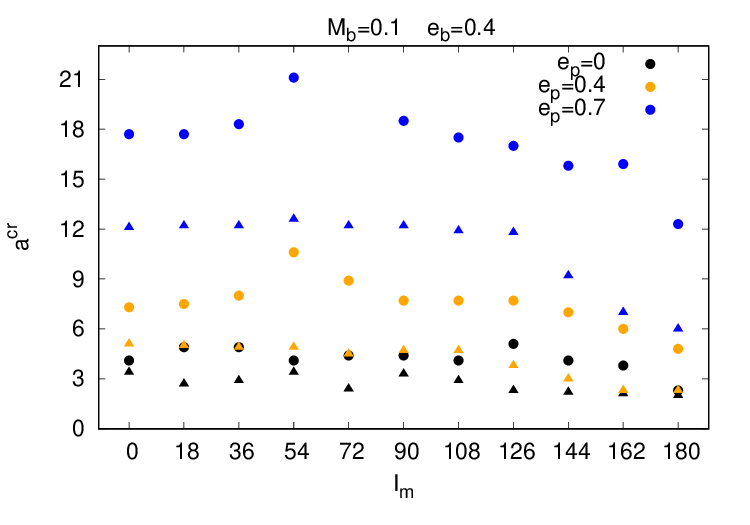}
\includegraphics[width=59mm,height=50mm]{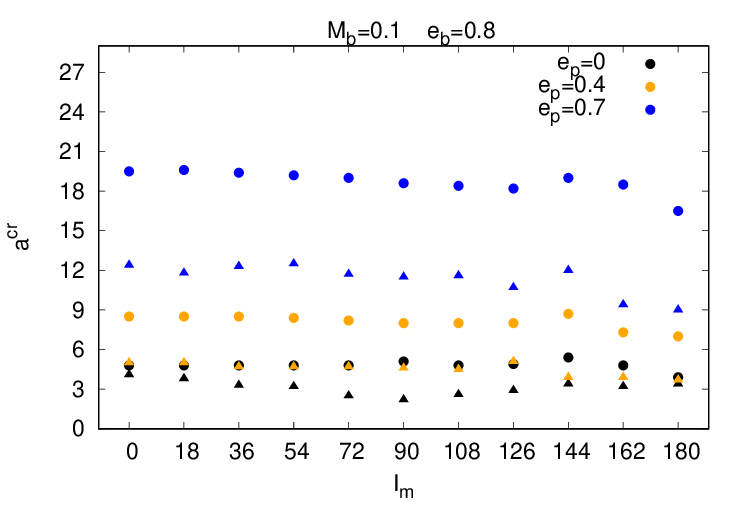}
\includegraphics[width=59mm,height=50mm]{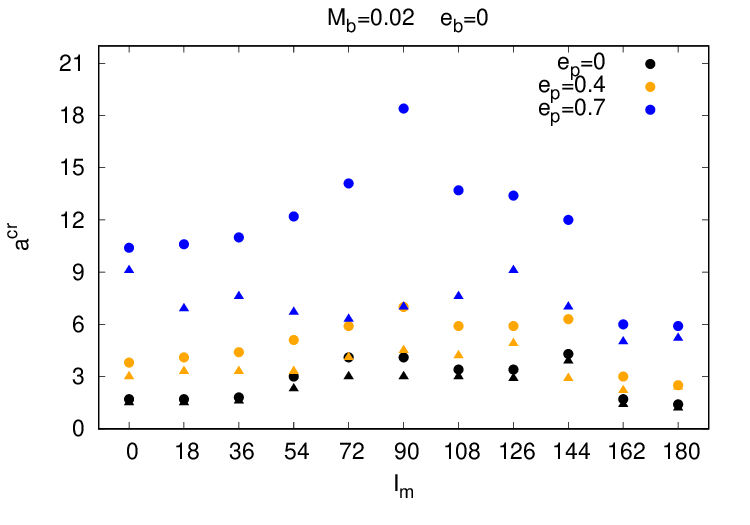}
\includegraphics[width=59mm,height=50mm]{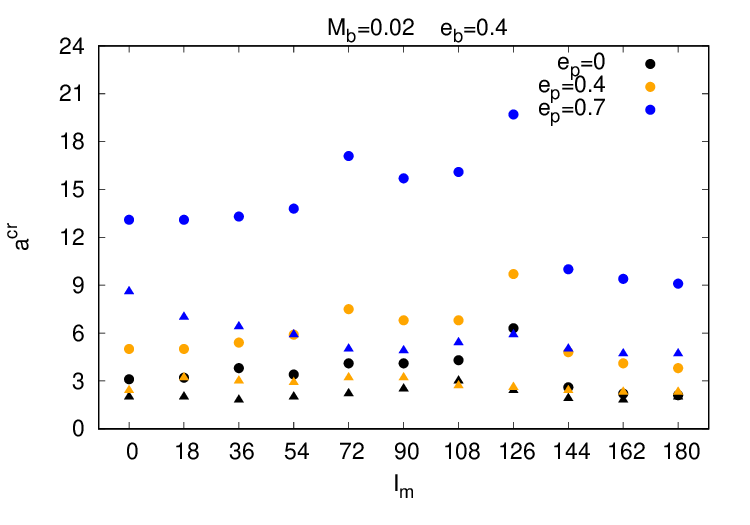}
\includegraphics[width=59mm,height=50mm]{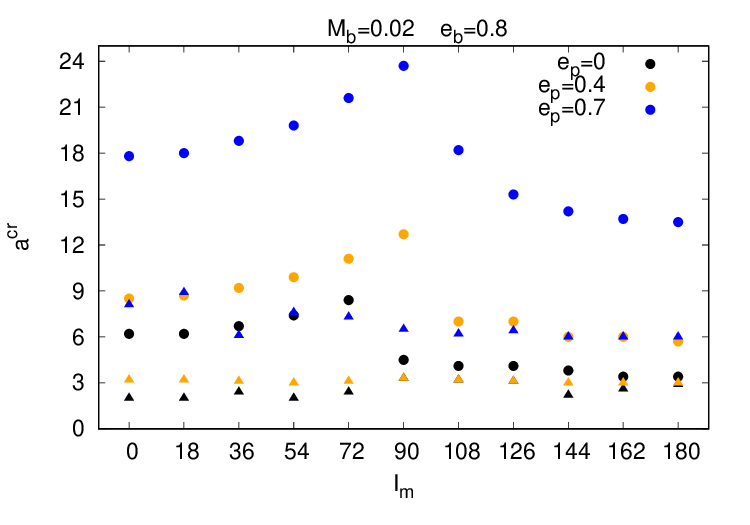}
\caption{Critical semi-major axis against mutual inclination for various eccentricity and binary mass ratio values. The circles denote the outer limit (largest value recorded among the different combinations of the planetary pericenter and node), while the inner limit is represented by triangles (smallest value recorded). The colors correspond to different planetary eccentricity values. Finally, $M_p=10^{-3}$.}
\label{fig7}
\end{center}
\end{figure}

Finally, figure \ref{fig7} contains a collection of plots covering a large portion of our parameter space and providing a sense of how the various parameter combinations affect the location of the critical semi-major axes.  The plots are another form of visualization of the findings discussed above.

\subsection{Fitting formulae}


In order to quantify the results of our numerical experiments and provide a tool for assessing the dynamical stability of a given system, we derived empirical formulae for the two critical borders over our entire parameter space. The procedure went as follows. For every specific combination of $M_b$, $M_p$, $I_m$, $e_b$ and $e_p$,  we simulated the dynamics of the system for nine initial relative starting positions of the binary and the planet corresponding to different pairs of  $\Omega_p - \omega_p$.  This resulted in nine inner critical planetary semi-major axes and nine outer critical semi-major axes for every set of $M_b$, $M_p$, $I_m$, $e_b$ and $e_p$ values. 
For a coplanar system with an initially eccentric planet or a non-coplanar system that had a planet on an initially circular orbit, the longitude of the ascending node in the former case and the argument of pericenter in the latter case, could not be defined.  Thus, we had only three inner and three outer critical semi-major axes values instead of nine for these particular types of systems.
 Finally, coplanar systems with circular planets only resulted in a unique value for each of the two critical borders. 
 
 When more than one critical semi-major-axis was available for a set of $M_b$, $M_p$, $I_m$, $e_b$ and $e_p$ parameters, we retained the largest value for the outer critical border and the smallest value for the inner critical border. Extreme values provide a better marker for the critical borders as defined in section (\ref{stability}).  At the same time, this reduces the number of independent variables by two, since $\Omega_p$ and $\omega_p$ are accounted for.  Moreover, we decided to drop the planetary mass from the fitting variables, since, as described in section \ref{sec:resu}, that specific parameter had little effect on the critical semi-major axes for circumbinary planets.

 In order to provide adequate fit performance for parameters that were allowed to vary by several orders of magnitude we re-scaled the binary mass ratio and the critical distances through a logarithm (base 10) while the mutual inclination was taken in radians. The remaining four independent variables, i.e. $M_b$, $I_m$, $e_b$ and $e_p$, were used in the fit, where polynomial models of the first, second and third order were tested against the data. A third order model in those variables best represented our dataset.  A $\chi^2$ goodness-of-fit test was used to confirm that our statistically derived empirical conditions reflected the simulation output. 

As a third order fit in four independent variables is somewhat unwieldy due to the large number of coefficients, we attempted to further reduce the complexity of the fit. 
We successively eliminated terms of the fit while monitoring the reduced $\chi^2$ parameter ($\chi_r^2=\chi^2/(n-1)$, n being the number of data points). After the decision of which terms to retain in the final model was made, we re-fitted the new reduced model to our data points in order to guarantee optimality of the coefficients in a least-squares sense. To enhance the quality of the overall result, we applied the above process on two data sets: the first one included planetary eccentricities of 0.8 or less while the second set included all values of the planetary eccentricity. This split was done in order to provide a better fit for the cases where $e_p \leq 0.8$ as the stability borders move considerably when $e_p=0.9$.

\subsection{Empirical fits for $e_p \leq 0.8$} \label{subsec:ep08}
If planetary eccentricity is confined to $e_p \leq 0.8$, the inner critical semi-major axis $a_i^{cr}$ and the outer critical semi-major axis $a_o^{cr}$ can be calculated using the following empirical fits: 

\begin{eqnarray}
a^{cr}_{i}&=& a_b \,\cdot 10^{[(\boldsymbol{B}_i\pm \boldsymbol{C}_i) \cdot \boldsymbol{X}_i]} \quad \text{and} \quad a^{cr}_{o}= a_b\,\cdot 10^{[(\boldsymbol{B}_o \pm \boldsymbol{C}_o)\cdot \boldsymbol{X}_o]} \label{fits},
\end{eqnarray}
where $(\boldsymbol{B}_{i,o} \pm \boldsymbol{C}_{i,o}) \cdot \boldsymbol{X}_{i,o}$ is the dot product between the fit coefficient vectors $\boldsymbol{B}$ (along with their uncertainty vectors $\boldsymbol{C}$) and the parameter vectors $\boldsymbol{X}$ for the inner and outer border respectively. With $M_{lb}=\log_{10}(M_b)$, the corresponding vectors read
\begin{eqnarray}
\begin{split}
\label{exp1}
\boldsymbol{B}_i &=& (0.20729,-0.32875,0.10339,0.58433,0.36623,-0.25569,-0.06425,-0.38387, \\
&& 1.01951,0.26910,0.38912,-0.19863,-0.25361,-0.30333,0.09080,-0.05955),\\
\boldsymbol{C}_i &=& (0.003763,0.01015,0.00224,0.00922,0.00978,0.00982,0.00069,0.00947, \\
&& 0.01176,0.00687,0.00759,0.00420,0.00735,0.00913,0.00129,0.00280),\\
\boldsymbol{X}_i &=& \left(1,M_{lb},I_m,e_b,e_p,M_{lb}^2,I_m^2,e_b^2,e_p^2,M_{lb}e_b,M_{lb}e_p,I_me_b,M_{lb}e_b^2,M_{lb}e_p^2,I_m^2e_b,M_{lb}^3\right).
\end{split}
\end{eqnarray}
For the outer critical planetary semi-major axis we find 
\begin{eqnarray}
\begin{split}
\boldsymbol{B}_o &=& (0.23612,-0.29377,0.22710,1.06753,0.62109,-0.21512,-0.06648,-1.52936,\\
& & -0.4748,-0.31329,-0.00869,0.11846,-0.03932,-0.00933,0.87506,1.25895),\\
\boldsymbol{C}_o &=& (0.00317,0.00927,0.00313,0.00905,0.00975,0.00910,0.00202,0.02330, \\	
&& 0.02893,0.00389,0.00116,0.00119,0.00260,0.00041,0.01699,0.02373),\\
\boldsymbol{X}_o &=& \left(1,\,M_{lb},\, I_m,\, e_b,\, e_p,\, M_{lb}^2,\, I_m^2,\, e_b^2,\, e_p^2,\, I_me_b,\, I_me_p,\, I_m^2e_b,\, M_{lb}^3,\, I_m^3, e_b^3,\, e_p^3\right)
\label{exp2}.
\end{split}
\end{eqnarray}

\subsection{Empirical fits for the entire dataset}\label{subsec:ep09}
Making use of the same structure of equations (\ref{fits}) for the inner and outer critical semi-major axes, we find the following coefficients and parameters that cover the entire dataset for the inner critical semi-major axis of the planet:
\begin{eqnarray}
\label{exp3}
\begin{split}
\boldsymbol{B}_i &=& (0.30889,-0.26446,0.09362,0.37426,0.31306,-0.27007,-0.06102,-0.09262,\\
& & 0.19436,-0.18911,-0.05466,0.06746,0.08715,1.19488), \\
\boldsymbol{C}_i &=& (0.00398,0.01150,0.00238,0.00688,0.00286,0.01061,0.00073,0.00472, \\
&& 0.00969,0.00447,0.00298,0.00417,0.00137,	0.00344),\\
\boldsymbol{X}_i &=& \left(1,M_{lb},I_m,e_b,e_p,M_{lb}^2,I_m^2,e_b^2,M_{lb}e_b,I_me_b,M_{lb}^3,M_{lb}^2e_b,I_m^2e_b,e_p^3  \right).
\end{split}
\end{eqnarray}
For the outer critical semi-major axis of the planet we have:
\begin{eqnarray}
\label{exp4}
\begin{split}
\boldsymbol{B}_o &=& (0.25556,-0.27038,0.20643,1.02175,0.80028,-0.21010,-0.08452,-1.46178,-1.20652,\\
& & -0.04965,-0.2989,-0.00227,-0.03860,0.01838,0.11341,0.83529,1.94189)\\
\boldsymbol{C}_o &=& (0.00324,0.00888,0.00285,0.00861,0.00844,0.00865,0.00086,0.02216, \\
&& 0.02216,0.00176,0.00370,0.0010,0.00247,0.00054,0.00113,0.01616,0.01616),\\
\boldsymbol{X}_o &=&\left(1,M_{lb},I_m, e_b,e_p,M_{lb}^2,I_m^2,e_b^2,e_p^2,M_{lb}I_m,I_me_b,I_me_p,M_{lb}^3,M_{lb}I_m^2,I_m^2e_b,e_b^3,e_p^3\right).
\end{split}
\end{eqnarray}

\begin{figure}
\begin{center}
\includegraphics[width=80mm,height=65mm]{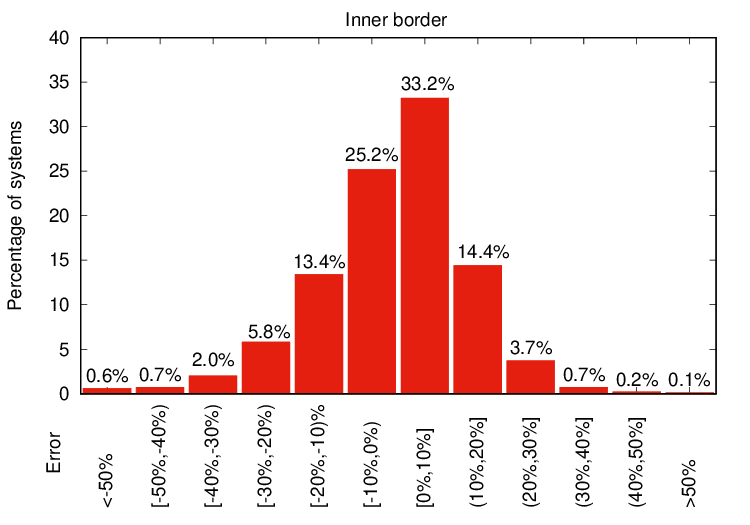}
\includegraphics[width=80mm,height=65mm]{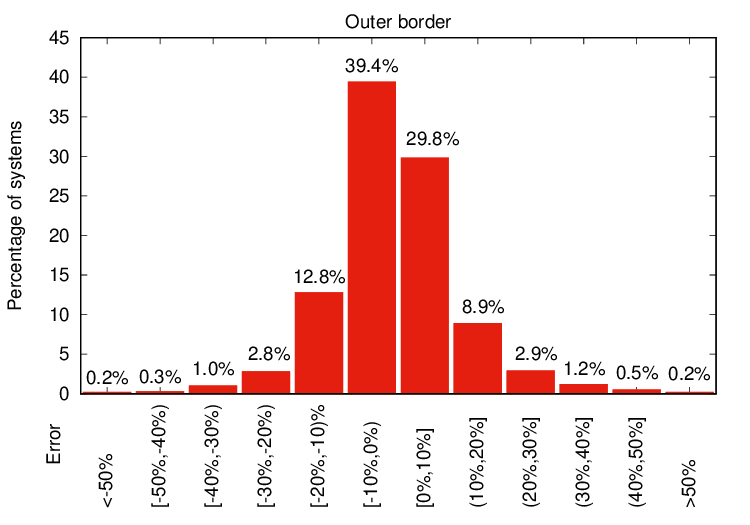}
\includegraphics[width=80mm,height=65mm]{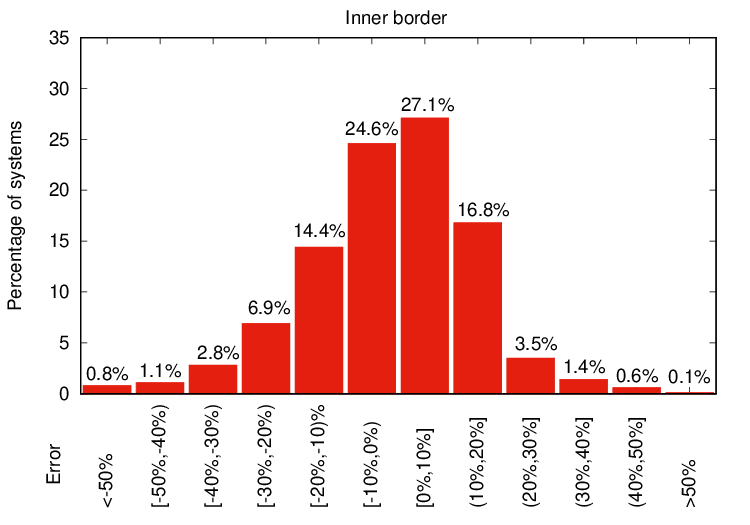}
\includegraphics[width=80mm,height=65mm]{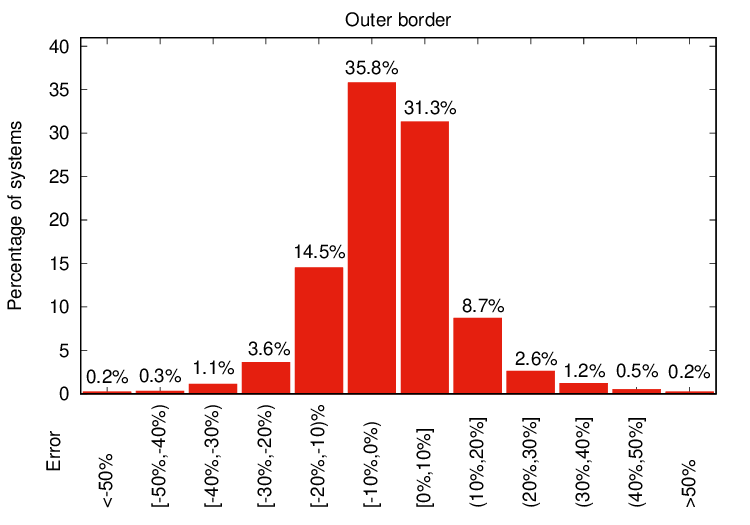}
\caption{Relative percentage error distribution from comparing our empirical fits against numerical simulation results. On the x-axis we have bins of relative percentage error between the results from the numerical simulations and the fits of equations  (\ref{fits}), while in the y-axis we have the percentage of systems that fall into a specific error bin.  The top row is for the $e_P \leq 0.8$ case, while the bottom row plots represent the more eccentric case.}
\label{fig8}
\end{center}
\end{figure}
\begin{figure}
\begin{center}
\includegraphics[width=80mm,height=60mm]{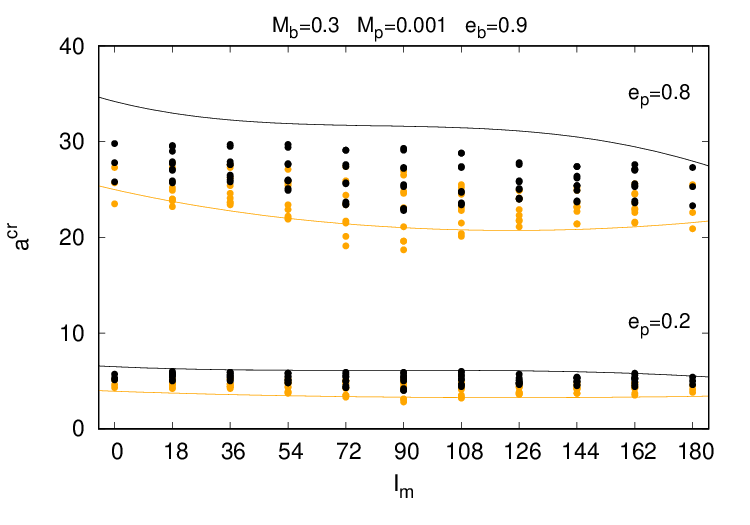}
\includegraphics[width=80mm,height=60mm]{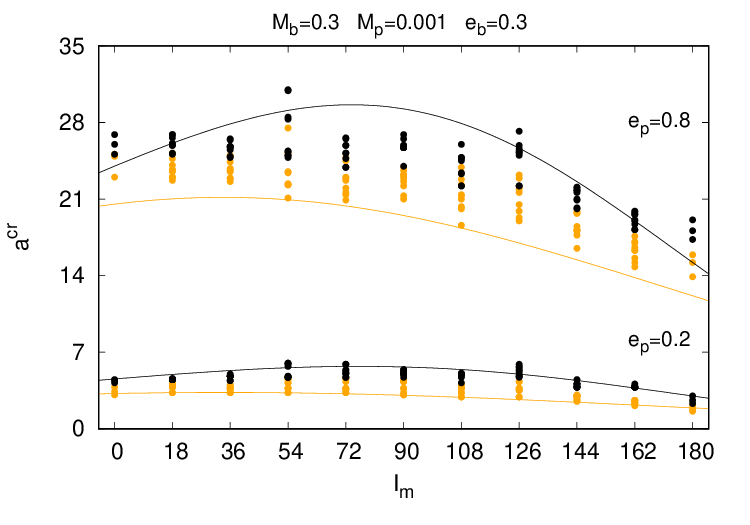}
\includegraphics[width=80mm,height=60mm]{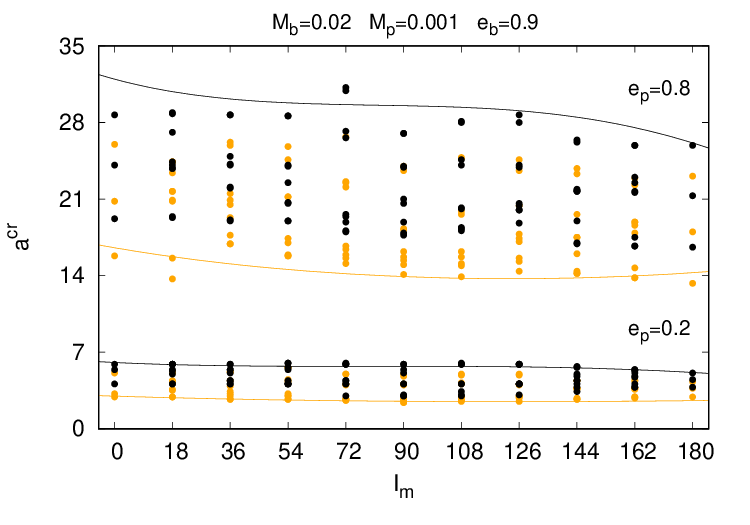}
\includegraphics[width=80mm,height=60mm]{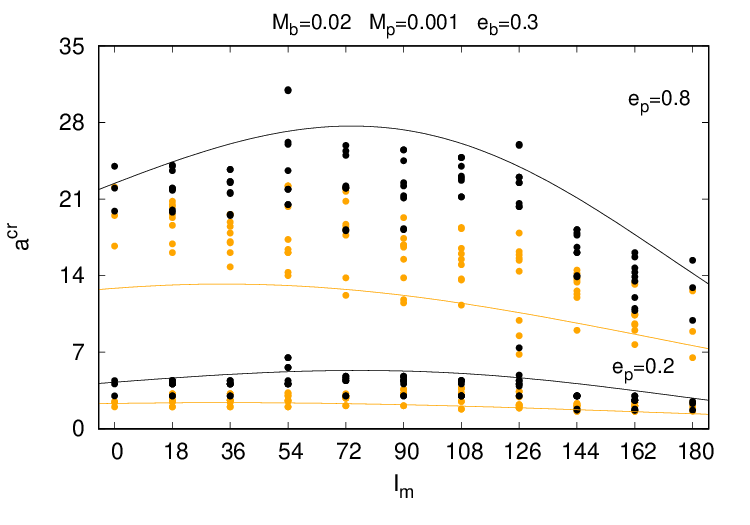}
\caption{Critical semi-major axis against mutual inclination for a variety of systems. The orange color refers to the inner boundary, while the black colour denotes the outer stability border. The continuous lines are our empirical fits as given in subsection \ref{subsec:ep08}. The points are the output from the numerical simulations for the specific systems. Note that the majority of the points lie between the two curves as they ideally should.} 
\label{fig9}
\end{center}
\end{figure}

A visual representation of the quality of the empirical fits can be found in Figures \ref{fig8} and \ref{fig9}.
In Figure \ref{fig8} we have a series of plots that show the distribution of the relative percentage error  
between the data points and the fits,  i.e. $100(a^{cr}_s-a^{cr}_{i,o})/a^{cr}_s, a^{cr}_s$ being the critical semi-major axis from the simulations.  In all cases, the majority of the errors 
are within $\pm10\%$. Errors larger than $50\%$ were encountered in less than two percent of all cases.  Figure \ref{fig9}   demonstrates the effectiveness of our empirical formulae for a sample of combinations of masses and orbital eccentricities.

\section{Fit performance against random simulations}\label{sec:fit_performance}

In order to test the quality of our fitting formulae, we carried out a number of additional, randomly generated, simulations.
The initial conditions for these simulations were created by using the pseudo-random number generating
GNU Fortran function $\it{rand}$.   The parameter values were drawn from a uniform distribution within the ranges given in section \ref{meth}, i.e.
$M_b \in [0.01,0.5]$, $e_b$ and $e_p$ $\in [0,0.9]$, $I_m \in [0^{\circ}, 180^{\circ}]$, $\varpi_p$, $\omega_p$ and $\Omega_p$ $\in [0^{\circ}, 360^{\circ}]$.  Regarding $M_p$, we drew values for $\log_{10}(M_p)$ uniformly  in the range
$[-7,-2]$. In order to sample the planetary semi-major axis, first, we created two distributions for that variable: one based on the dataset for the inner critical semi-major axis and one based on the dataset for the outer critical semi-major axis. Then,  we used rejection sampling to create a set of 25,000 random systems based on the inner critical value distribution and another set of equal number of systems using the outer critical value distribution.  The random seed number for the first set was 121, while for the other one the seed was 446.


The results of the random simulations and how they were classified based on our fitting formulae are presented in Tables (\ref{tablerandom1}) and (\ref{tablerandom2}).
As stated in section \ref{meth}, the planet is initiated at eight different positions around the binary, while the binary was started either at pericenter or at apocenter.
That makes a total of sixteen initial positions for our systems. By definition, we would like to avoid encountering unstable orbits beyond the outer critical semi-major axis
and any stable orbits below the inner critical semi-major axis.  Therefore, we define our criteria for success as follows: the fits are predicting the stability of a system successfully when systems with 0 unstable positions were found in the stable zone, systems with 16 unstable positions were found in the unstable zone and finally, when all other cases were in between those two areas.

Generally, the outcome of the comparison between the fits and the results of the random system simulations was very good. If we consider the fits for $e_p \le 0.8$, the majority of the fully stable and fully unstable systems was found in the right place. Only 1 case of a $0$ and 54 cases of $16$s were found in the wrong area (i.e.  0s in the fully unstable and 16s in the fully stable area).  Also,  $88.0\%$ of the systems with mixed stability behavior were correctly classified. 
In total, the success rate was $80.4\%$. The above mentioned numbers refer to the random systems that were drawn from the inner border distribution, but the pattern was very similar for the systems drawn from the outer border distribution.  The success rate for those systems was $83.7\%$.
The classification of the random systems for the restricted dataset can be found in Table \ref{tablerandom1}, where the numbers in bold at the diagonal positions of the table, when added up, provide the total success rate numbers mentioned above.

These percentages, however, are conservative estimates. This is because, as we saw with the results of the original set of simulations,
a fully stable or a fully unstable system, at a specific semi-major axis, may occasionally appear in the mixed zone. In that case, we can add to the previous percentages the 0s and the 16s that are found in the mixed zone (the numbers in bold in Table \ref{tablerandom1} that are not lying on its diagonal).  When we do that, the $80.4\%$ and $83.7\%$ rise to $98.7\%$ and $98.9\%$ respectively.  These numbers now represent the best case scenario. Nonetheless, we expect that the success rate would be somewhere between those extreme values as we do not know the exact location of the 0s and 16s since for a specific set of masses, inclination, eccentricities and slow angles we only have results for a specific semi-major axis and
not a whole range of them, contrary to what the case was with our initial simulations that were used to derive the fits.  As seen in Table \ref{tablerandom2}, where the random simulation data are classified using fits to the entire dataset, the success prediction rates are very similar to those obtained in the case of the fits for $e_p\le0.8$.

\begin{table}
\begin{center}	
\caption{Random simulation result classification using the fits given in section \ref{subsec:ep08}.} 
\label{tablerandom1}
\vspace{0.1 cm}
{\begin{tabular}{c c c c c c}
\hline\hline\\
          &          &           &    INNER BORDER DISTRIBUTION  & \\
\vspace{0.2cm}\\          
\hline
          &          &           &     Quantity per zone      & \\
 N.U.I.P. & & Stable & Mixed & Unstable & Total \\
\hline
$0$ & &${\bf 12148 (54.7\%)}$  &  ${\bf 1826 (8.2\%)}$ & $ 1 (0.0\%)$ & $ 13975 (62.9\%)$ \\
$1-15$ & &$ 216 (1.0\%)$  & $ {\bf 1825 (8.2\%)}$ & $ 32 (0.1\%)$ & $ 2073 (9.3\%)$ \\
$16$ & &$ 54 (0.2\%)$     &  ${\bf 2245 (10.1\%)}$  & $ {\bf 3877 (17.5\%)}$ & $ 6176 (27.8\%)$ \\
\hline
& Total & $ 12418 (55.9\%)$  & $ 5896 (26.5\%)$  & $ 3910 (17.6\%)$ & $ 22224 (100.0\%)$ \\
\hline\\
\vspace{0.2cm}
          &          &           &    OUTER BORDER DISTRIBUTION  & \\
\vspace{0.2cm}\\          
\hline
          &          &           &     Quantity per zone       & \\
 N.U.I.P. & & Stable & Mixed & Unstable & Total \\
\hline
$0$ & & ${\bf 14905 (66.8\%)}$  & $ {\bf 1712 (7.7\%)}$ & $ 2 (0.0\%)$ & $ 16619 (74.5\%)$ \\
$1-15$ & &$ 177 (0.8\%)$  & $ {\bf 1468 (6.5\%)}$ & $ 18 (0.1\%)$ & $ 1663 (7.4\%)$ \\
$16$ & &$ 37 (0.2\%)$  &  ${\bf 1685 (7.5\%)}$  & $ {\bf 2320 (10.4\%)}$ & $ 4042 (18.1\%)$ \\
\hline
& Total & $ 15119 (67.8\%)$  & $ 4865 (21.8\%)$  & $ 2340 (10.5\%)$ & $ 22324 (100.0\%)$ \\
\hline
\end{tabular}}
\end{center}
\tablecomments{N.U.I.P: Number of unstable initial positions of the planet on its initial orbit. The percentages refer to the number of individual cases over the total number of cases.  For the interpretation of numbers in bold please see the main body of the paper.}
\end{table}

\begin{table}
\begin{center}	
\caption{Random simulation result classification using the fits given in section \ref{subsec:ep09}.} 
\label{tablerandom2}
\vspace{0.1 cm}
{\begin{tabular}{c c c c c c}
\hline\hline\\
          &          &           &    INNER BORDER DISTRIBUTION  & \\
\vspace{0.2cm}\\          
\hline
          &          &           &     Quantity per zone      & \\
 N.U.I.P. & & Stable & Mixed & Unstable & Total \\
\hline
$0$ & &${\bf 13268 (53.1\%)}$  &  $ {\bf  2009 (8.0\%)}$ & $ 2 (0.0\%)$ & $ 15279 (61.1\%)$ \\
$1-15$ & &$ 231 (0.9\%)$  & $ {\bf 1955 (7.8\%)}$ & $ 90 (0.4\%)$ & $ 2276 (9.1\%)$ \\
$16$ & &$ 66 (0.3\%)$  &  $ {\bf  2513 (10.1\%)}$  & $ {\bf 4866 (19.4\%)}$ & $ 7445 (29.8\%)$ \\
\hline
& Total & $ 13565 (54.3\%)$  & $ 6477 (25.9\%)$  & $ 4958 (19.8\%)$ & $ 25000 (100.0\%)$ \\
\hline\\
\vspace{0.2cm}
          &          &           &    OUTER BORDER DISTRIBUTION  & \\
\vspace{0.2cm}\\          
\hline
          &          &           &     Quantity per zone       & \\
 N.U.I.P. & & Stable & Mixed & Unstable & Total \\
\hline
$0$ & &${\bf 16294 (65.2\%)}$  &  $ {\bf  1791 (7.1\%)}$ & $ 2 (0.0\%)$ & $ 18087 (72.3\%)$ \\
$1-15$ & &$ 183 (0.7\%)$  & $ {\bf 1552 (6.2\%)}$ & $ 57 (0.3\%)$ & $ 1792 (7.2\%)$ \\
$16$ & & $38 (0.1\%)$  & ${\bf 2014 (8.1\%)}$  & $ {\bf 3069 (12.3\%)}$ & $ 5121 (20.5\%)$ \\
\hline
& Total & $ 16515 (66.0\%)$  & $ 5357 (21.4\%)$  & $ 3128 (12.6\%)$ & $ 25000 (100.0\%)$ \\
\hline
\end{tabular}}
\end{center}
\tablecomments{The notation, numbers and colors have the same meaning as in Table \ref{tablerandom1}.}
\end{table}

\section{Empirical Fitting with Machine Learning}\label{sec:ml}

In addition to empirical formulae derived via least squares fits to our numerical dataset, we also explored models that can account for higher order correlations and additional non-linearities by using Machine Learning (ML). The ML approach serves two purposes: i) benchmarking our analytical fits against another technique, and ii) providing a simple, trained model as an alternative that does not require manual implementation of the empirical formulae.

We trained and tested an \texttt{XGBRegressor} model \citep{xgboost}, which is a type of gradient boosted trees algorithm, using its default hyperparameters (settings that control the model's learning process) on the same data set used for the empirical fits. Gradient boosted trees is a learning method that combines a large number of decision trees (simple ``weak learner'' models) into an ``ensemble'' to create a more accurate model (a ``strong learner''). This approach has many advantages, including high predictive accuracy, resistance to data noise, and simplicity of implementation.

\begin{figure}
\begin{center}
\includegraphics[width=80mm,height=60mm]{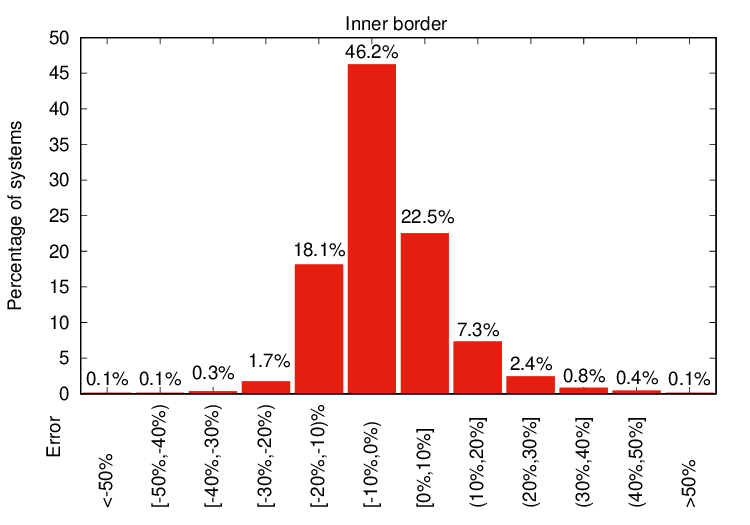}
\includegraphics[width=80mm,height=60mm]{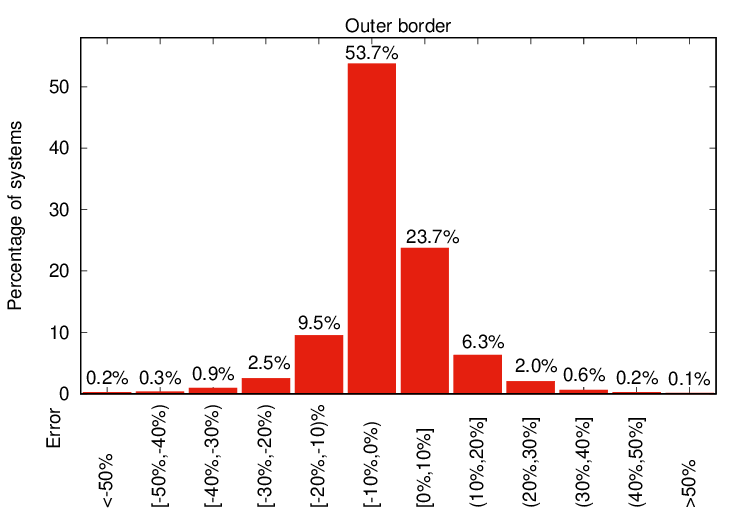}
\includegraphics[width=80mm,height=60mm]{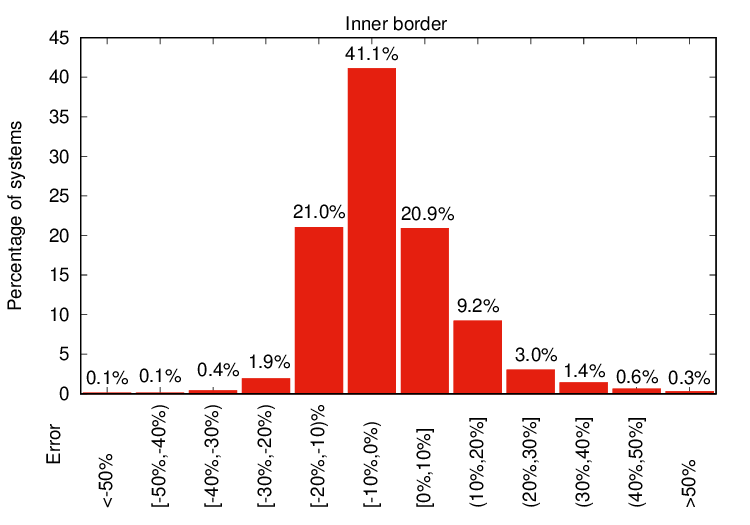}
\includegraphics[width=80mm,height=60mm]{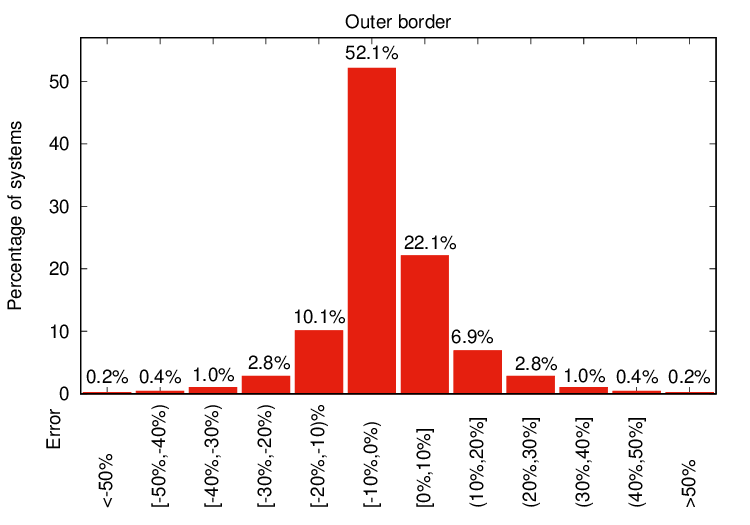}
\caption{Relative percentage error distribution from comparing the Machine Learning model against the results from the numerical simulations. On the x-axis we have bins of relative percentage error between the Machine Learning model and the results from the numerical simulations, while in the y-axis we have the percentage of systems that fall into a specific error bin.  The top row corresponds to systems with $e_p \leq 0.8$ case, while the bottom row plots represent the complete dataset.}
\label{fig10}
\end{center}
\end{figure}

Similar to  our approach for generating the empirical fits for the stability limits, we trained two separate models: one that predicts stability limits for planetary eccentricities $e_p \leq 0.8$ and one for $e_p \leq 0.9$. The goodness-of-fit of the models was initially evaluated through the mean absolute error of 10-fold cross-validation\footnote{K-fold cross-validation is a technique where the dataset is divided into k equal parts and the model is trained and tested k times, each time using a different part for testing and the remaining parts for training.} with the models trying to predict the inner and outer borders. The resulting goodness-of-fit values  ranged from $\sim$ 0.05 for the circular cases, to $\sim$ 0.3 for the eccentric cases, although those are skewed by the Mean Absolute Error (MAE) sensitivity to outliers. There were no significant differences for any given case between the MAE of the inner and outer borders. Moreover, a grid search over hyperparameters\footnote{exhaustively searching through a predefined set of hyperparameter combinations to find the optimal configuration for a machine learning model.} and testing other ML algorithms such as \texttt{LightGBM} and \texttt{Random Forest Regressor} did not meaningfully improve the results.  Figure \ref{fig10} illustrates the performance of the ML models on the training datasets. The plots in this figure show that the errors of the Machine Learning models are more tightly clustered towards the center of the bar charts compared to the errors of the empirical fits presented in Figure \ref{fig8}, which implies that the Machine Learning models are better at capturing outliers.

Similar to our empirical formulae, we tested the Machine Learning models against the results from the random simulations we conducted. The outcomes of this comparison are presented in Tables \ref{tablerandom3} and \ref{tablerandom4}. Overall, the results are consistent with and resemble those from the comparison between the empirical formulae and the random simulations, although some differences are noted. First, there are more systems, from all categories,  classified above the outer critical limit; about up to $2\%$ more than the empirical fits. Second, there are more ``16s'' in the zone between the two limits (of the order of $ 1\% $). These two differences imply that we have fewer 16s below the lower critical limit.  Finally, in all but one case, we recorded more fully stable systems in the fully unstable area. But those cases were rare. Regarding the success rates, these seem to be similar to those scored by the empirical fits. The empirical formulae seem to do just a bit better ($1\%$) in most of the cases. Hence, both approaches result in having useful tools for characterizing the stability of circumbinary planets and may be used according to preference.  Our trained machine learning models and a simple inference code can be found in \cite{mlmodel}.



\begin{table}
\begin{center}	
\caption{Random simulation result classification using the Machine Learning model ($e_p \leq 0.8$).} 
\label{tablerandom3}
\vspace{0.1 cm}
{\begin{tabular}{c c c c c c}
\hline\hline\\
          &          &           &    INNER BORDER DISTRIBUTION  & \\
\vspace{0.2cm}\\          
\hline
          &          &           &     Quantity per zone      & \\
 N.U.I.P. & & Stable & Mixed & Unstable & Total \\
\hline
$0$ & &${\bf 12658 (57.0\%)}$  &  $ {\bf  1312 (5.9\%)}$ & $ 5 (0.0\%)$ & $ 13975 (62.9\%)$ \\
$1-15$ & &$ 382 (1.7\%)$  & $ {\bf 1671 (7.5\%)}$ & $ 20 (0.1\%)$ & $ 2073 (9.3\%)$ \\
$16$ & &$ 133 (0.6\%)$     &  $ {\bf 2560 (11.5\%)}$  & $ {\bf 3483 (15.7\%)}$ & $ 6176 (27.8\%)$ \\
\hline
& Total & $ 13173 (59.3\%)$  & $ 5543 (24.9\%)$  & $ 3508 (15.8\%)$ & $ 22224 (100.0\%)$ \\
\hline\\
\vspace{0.2cm}
          &          &           &    OUTER BORDER DISTRIBUTION  & \\
\vspace{0.2cm}\\          
\hline
          &          &           &     Quantity per zone       & \\
 N.U.I.P. & & Stable & Mixed & Unstable & Total \\
\hline
$0$ & &${\bf 15415 (69.1\%)}$  & $ {\bf 1203 (5.4\%)}$ & $ 1 (0.0\%)$ & $ 16619 (74.5\%)$ \\
$1-15$ & &$ 300 (1.3\%)$  & $ {\bf 1344 (6.0\%)}$ & $ 19 (0.1\%)$ & $ 1663 (7.4\%)$ \\
$16$ & &$ 99 (0.5\%)$  &  $ {\bf 1884 (8.4\%)}$  & $ {\bf 2059 (9.2\%)}$ & $ 4042 (18.1\%)$ \\
\hline
& Total & $ 15814 (70.9\%)$  & $ 4431 (19.8\%)$  & $ 2079 (9.3\%)$ & $ 22324 (100.0\%)$ \\
\hline
\end{tabular}}
\end{center}
\tablecomments{The notation, numbers and colors have the same meaning as in Tables \ref{tablerandom1} and \ref{tablerandom2}.}
\end{table}

\begin{table}
\begin{center}	
\caption{Random simulation result classification using the Machine Learning model ($e_p \leq 0.9$).} 
\label{tablerandom4}
\vspace{0.1 cm}
{\begin{tabular}{c c c c c c}
\hline\hline\\
          &          &           &    INNER BORDER DISTRIBUTION  & \\
\vspace{0.2cm}\\          
\hline
          &          &           &     Quantity per zone      & \\
 N.U.I.P. & & Stable & Mixed & Unstable & Total \\
\hline
$0$ & &${\bf 13630 (54.5\%)}$  &  $ {\bf  1616 (6.5\%)}$ & $ 33 (0.1\%)$ & $ 15279 (61.1\%)$ \\
$1-15$ & &$ 435 (1.8\%)$  & $ {\bf 1786 (7.1\%)}$ & $ 55 (0.2\%)$ & $ 2276 (9.1\%)$ \\
$16$ & &$ 182 (0.7\%)$  &  $ {\bf  2833 (11.4\%)}$  & $ {\bf 4430 (17.7\%)}$ & $ 7445 (29.8\%)$ \\
\hline
& Total & $ 14247 (57.0\%)$  & $ 6235 (25.0\%)$  & $ 4518 (18.0\%)$ & $ 25000 (100.0\%)$ \\
\hline\\
\vspace{0.2cm}
          &          &           &    OUTER BORDER DISTRIBUTION  & \\
\vspace{0.2cm}\\          
\hline
          &          &           &     Quantity per zone       & \\
 N.U.I.P. & & Stable & Mixed & Unstable & Total \\
\hline
$0$ & &${\bf 16592 (66.3\%)}$  &  $ {\bf  1488 (6.0\%)}$ & $ 7 (0.0\%)$ & $ 18087 (72.3\%)$ \\
$1-15$ & &$ 328 (1.3\%)$  & $ {\bf 1428 (5.7\%)}$ & $ 36 (0.2\%)$ & $ 1792 (7.2\%)$ \\
$16$ & &$ 166 (0.7\%)$  & $ {\bf 2131 (8.5\%)}$  & $ {\bf 2824 (11.3\%)}$ & $ 5121 (20.5\%)$ \\
\hline
& Total & $ 17086 (68.3\%)$  & $ 5047 (20.2\%)$  & $ 2867 (11.5\%)$ & $ 25000 (100.0\%)$ \\
\hline
\end{tabular}}
\end{center}
\tablecomments{The notation, numbers and colors have the same meaning as in our previous Tables.}
\end{table}

\section{Application to known circumbinary planetary systems.}

Among the exoplanets that have been discovered as of today, a number resides in circumbinary orbits. In this section we apply our stability criterion to the Kepler and Tess circumbinary systems that are currently known, i.e. Kepler-16 \citep{2011Sci...333.1602D}, Kepler-34 and Kepler-35 \citep{2012Natur.481..475W}, Kepler-38 \citep{2012ApJ...758...87O}, Kepler-47 \citep{2012Sci...337.1511O,2019AJ....157..174O}, Kepler-64 \citep{2013ApJ...768..127S, 2013ApJ...770...52K}, Kepler-413 \citep{2014ApJ...784...14K}, Kepler-453 \citep{2015ApJ...809...26W}, Kepler-1647 \citep{2016ApJ...827...86K}, Kepler-1661 \citep{2020AJ....159...94S}, TIC 172900988 \citep{2021AJ....162..234K} and TOI-1338 \citep{2023NatAs...7..702S}.  In order to validate our predictions  with the above systems we make use of the parameters given in the respective discovery papers. These parameter values can be found in Table \ref{kepler}.  

Evaluating our empirical fits for the above systems, we find that all but one planet are beyond the outer critical semi-major axis.  The values of the critical axes can be found in Table \ref{kepler1}.  Kepler-34 is the only exception with its outer critical semi-major axis being 1.092 au, while the planet is located at 1.090 au.  Generally, these results are in agreement with stability analyses in the discovery papers and other investigations specifically focused on some of those circumbinary systems \citep[e.g.][]{2015MNRAS.446.1283C,2016AstL...42..474P}. 

We would like to point out, that, for this cursory analysis we neglected planet - planet interactions for the two multi-planet systems Kepler-47 and TOI-1338, i.e. we treated them as circumbinary systems having one planet at a time. Regarding Kepler-47, \cite{2012Sci...337.1511O,2019AJ....157..174O} (and references therein) state that the unstable area around the binary extends around 0.18 au which is in good agreement with what we have found. For TOI-1338c, as the inclination and the longitude of the node of the outer planet are not known, \cite{2023NatAs...7..702S} found that the system gets unstable for mutual inclinations $40^{\circ} < I_m < 120^{\circ}$. On the other hand, if we vary the mutual inclination in our fits, we note that the outer planet always lies within the stable zone. 

\begin{table}
\begin{center}	
\caption{ Circumbinary system parameter values used for the validation of the empirical stability fits.} \label{kepler}
\vspace{0.1 cm}
{\begin{tabular}{c c c c c c c c c}\hline\hline\\
System & $m_1 (M_{\odot})$ & $m_2 (M_{\odot})$ & $m_p (M_J) $ & $I_m (^{\circ})$ & $a_b (au) $ & $a_p (au)$ & $e_b$ & $e_p$ \\
\hline
 Kepler-16  & 0.6897 & 0.20255 & 0.333 & 0.4 & 0.22431 & 0.7048 & 0.15944 & 0.00685 \\
 Kepler-34  & 1.0479 & 1.0208 & 0.22 & 1.81 & 0.22882 & 1.0896 & 0.52087 & 0.182 \\
 Kepler-35  & 0.8876 & 0.8094 & 0.127 & 1.28 & 0.17617 & 0.60345 & 0.1421 & 0.042 \\
 Kepler-38  & 0.949 & 0.249 & 0.384 & 0.182 & 0.1469 & 0.4646 & 0.1032 & 0.032 \\
 Kepler-47b & 0.957 & 0.342 & 0.006513 & 0.166 & 0.08145 & 0.2877 & 0.0288 & 0.021 \\
 Kepler-47c & 0.957 & 0.342 & 0.05984 & 1.165 & 0.08145 & 0.6992 & 0.0288 & 0.024 \\
 Kepler-47d & 0.957 & 0.342 & 0.00997 & 1.38 & 0.08145 & 0.9638 & 0.0288 & 0.044 \\
 Kepler-64  & 1.528 & 0.378 & 0.531 & 2.814 & 0.1744 & 0.634 & 0.2117 & 0.0539 \\
 Kepler-413  & 0.82 & 0.5423 & 0.21 & 4.073 & 0.10148 & 0.3553 & 0.0365 & 0.1181 \\
 Kepler-453  & 0.944 & 0.1951 & 0.05 & 2.258 & 0.18539 & 0.7903 & 0.0524 & 0.0359 \\
 Kepler-1647 & 1.21 & 0.975 & 1.52 & 2.9855 & 0.1276 & 2.7205 & 0.1593 & 0.0581 \\
 Kepler-1661 & 0.841 & 0.262 & 0.053 & 0.93 & 0.187 & 0.633 & 0.112 & 0.057 \\
 TIC 172900988 & 1.2388 & 1.2023 & 2.74 & 1.45 & 0.191928 & 0.89809 & 0.44793 & 0.088 \\
 TOI-1338b & 1.127 & 0.3313 & 0.0685 & 0 & 0.1321 & 0.4607 & 0.155522 & 0.088 \\
 TOI-1338c & 1.127 & 0.3313 & 0.205 & 0-180 & 0.1321 & 0.794 & 0.155522 & 0.16 \\
 
\hline
\end{tabular}}
\end{center}
\end{table}

\begin{table}
\begin{center}	
\caption{Stability borders for known circumbinary systems.} \label{kepler1}
\vspace{0.1 cm}
{\begin{tabular}{c c c c}\hline\hline\\
System & $a^{cr}_i (au)$ & $a^{cr}_o (au)$ & $a_p (au)$ \\
\hline
 Kepler-16  & 0.551 & 0.688 & 0.705  \\
 Kepler-34  & 0.804 & 1.092 & 1.090  \\
 Kepler-35  & 0.410 & 0.511 & 0.603 \\
 Kepler-38  & 0.349 & 0.427 & 0.464 \\
 Kepler-47 (b) & 0.178 & 0.198 & 0.288\\
 Kepler-47 (c) & 0.179 & 0.200 & 0.699 \\
 Kepler-47 (d) & 0.181 & 0.206 & 0.964 \\
 Kepler-64  & 0.457 & 0.627 & 0.634 \\
 Kepler-413  & 0.236 & 0.281 & 0.355 \\
 Kepler-453  & 0.427 & 0.504 & 0.790 \\
 Kepler-1647 & 0.310 & 0.397 & 2.720 \\
 Kepler-1661 & 0.452 & 0.570 & 0.633 \\
 TIC 172900988 & 0.579 & 0.784 & 0.898 \\
 TOI-1338 (b) & 0.337 & 0.448 & 0.461 \\
 TOI-1338  (c)& 0.361 & 0.492 & 0.794 \\
\hline
\end{tabular}}
\end{center}
\end{table}

\section{Online portal and API} \label{sec:online}
This section introduces a reboot of ExoStab \citep{pilat2011exostab}, the Exostab 2.0 Application Programming Interface (hereafter referred to as the API)\footnote{\url{https://exostab2.readthedocs.io/latest/}}. Exostab 2.0 \citep{exostab2} is a software interface designed to facilitate interaction with large catalogs of numerical stability simulations such as constructed in this work. The API enables external programs and applications to access and utilize the functionalities of the stability catalog in a controlled and standardized manner. The API follows a RESTful architectural style, providing a well-defined set of endpoints for data retrieval, manipulation, and control. In particular, stability limits for system configurations can be queried for and the closest matches in our database are returned in JSON format. The reasonably dense catalog grid makes this process meaningful. The real advantage of returning nearest neighbors is the circumnavigation of issues that arise in fitting and interpolation, such as the mixing of dynamical behavior of systems in and out of resonance, etc. Only stability limits that were actually calculated are being returned. 
A web-based front-end\footnote{\url{https://apexgroup.web.illinois.edu/stability/index.html}} allows for convenient visualization and access to query results in a formatted table as well as through as CSV file download. 

\section{Discussion} 
We have seen in sections \ref{sec:fit_performance} and \ref{sec:ml} that classifiers for dynamical stability are not perfect regardless of whether they were constructed empirically or through Machine Learning.  
As such tools are nevertheless used frequently in the community, we now proceed to compare our results with other, popular stability criteria, in order to put the limitations of our models into context.

\subsection{Comparison with other results}
\begin{figure}
\begin{center}
\includegraphics[width=80mm,height=65mm]{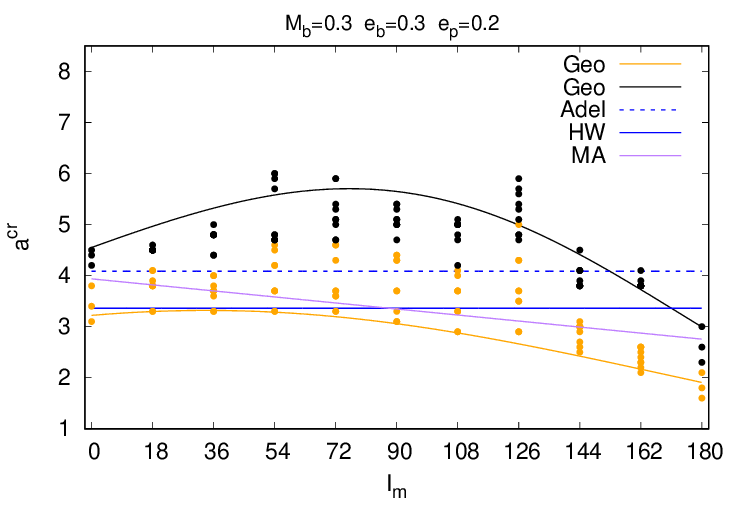}
\includegraphics[width=80mm,height=65mm]{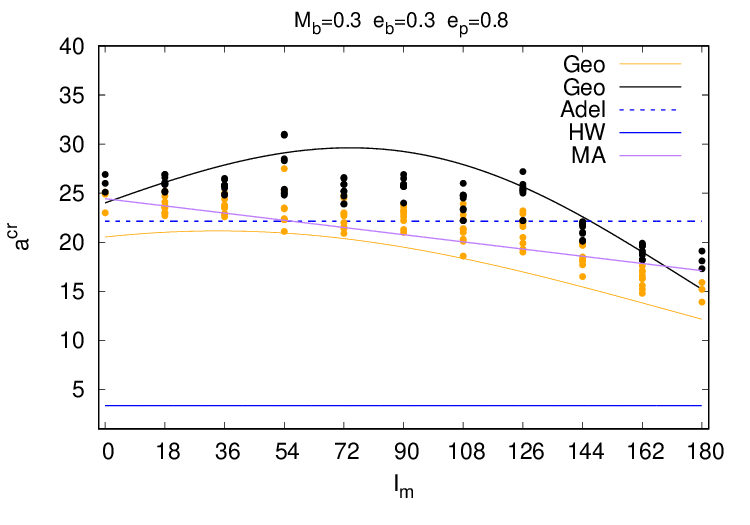}
\includegraphics[width=80mm,height=65mm]{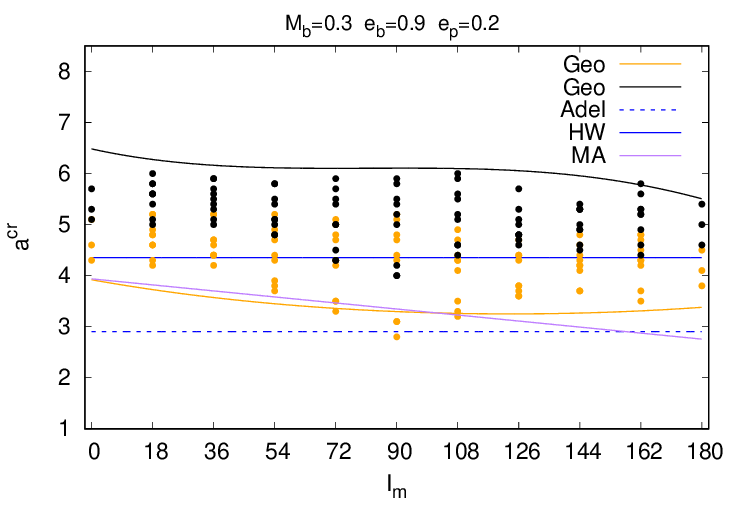}
\includegraphics[width=80mm,height=65mm]{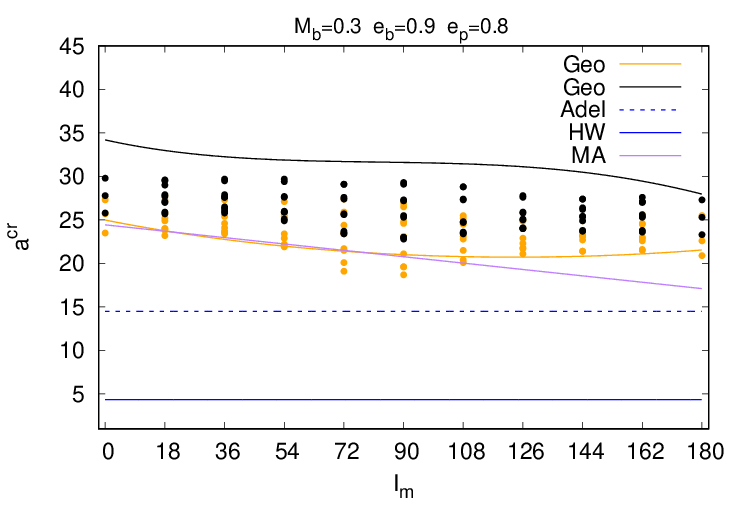}
\caption{Comparison between different stability fits. Geo stands for the fits of the present work, Adel denotes the work by \cite{2023A&A...680A..29A},
 HW denotes the classification formula given in \cite{1999AJ....117..621H} and MA stands for the criterion developed by \cite{1999ASIC..522..385M,2001MNRAS.321..398M} . As previously, the black color denotes the outer critical border, while the orange color represents the inner critical border. The circles are the results from our numerical simulations.}
\label{fig11}
\end{center}
\end{figure}

Figure \ref{fig11} contains a graphical comparison of various stability criteria, including those developed in this study,
for systems with inclined and eccentric orbits. The empirical fits presented in this work fare substantially better than \citet{1999AJ....117..621H}, \cite{2001MNRAS.321..398M} and \cite{2023A&A...680A..29A} in capturing the dynamical behavior of such configurations. The stability criterion by \cite{1999AJ....117..621H}, for instance, does not account for planetary eccentricity, and thus, fails to correctly predict where the transition between stable and unstable orbits takes place. The formula of \cite{2023A&A...680A..29A} does better for eccentric planetary orbits compared to \cite{1999AJ....117..621H}, but for binary stars on highly eccentric orbits, predictions of dynamical stability based on equation (\ref{adeleq}) are poor. Of course, both the criteria in \cite{1999AJ....117..621H} and in \cite{2023A&A...680A..29A} are derived from and for coplanar systems, which is another substantial limitation. On the other hand, although the criterion of \cite{2001MNRAS.321..398M} applies to three dimensional orbits, its linear dependence on $I_m$ clearly fails to capture the variations of the stability border due to the mutual inclination.  This is to be expected, as the non-linear dependence of the stability limit on the mutual inclination has been noticed before \citep[e.g.][]{2013NewA...23...41G,2022MNRAS.516.4146V,2022PASA...39...62T}.  In addition, equation (\ref{marda}) is not a function of $M_b$ and $e_b$. These two parameters, however, have a moderate effect on determining the stability borders as discussed in section  \ref{sec:resu} and seen in figure \ref{fig2}. Finally, the stability curve given by equation (\ref{marda}) appears to be much closer to our inner critical border than to the outer one and therefore many unstable orbits would be classified as stable.

\subsection{Non-Newtonian effects}\label{sec:rel}


Effects other than Newtonian gravity can be of importance for the orbital evolution of hierarchical triple systems. Examples range from the relativistic precession of the binary pericenter over tidal friction between the two stars to the deformation of the shape of the stars due to tides and rotation \citep[e.g.][]{2007ApJ...669.1298F,2012ApJ...747....4P,Naoz_2013,2015MNRAS.447..747L,2016CeMDA.126..189C}.  

Tidal friction between the stars may eventually shrink their orbit and reduce eccentricity. Depending on the distance between the stars, the timescale for any orbital changes that may eventually affect the planetary orbit can be quite long.  For instance, the eccentricity and semi-major axis of the stars in Kepler-34 were reduced by about $5\%$ and $3\%$ respectively after 8 Gyrs \citep{2015MNRAS.446.1283C}. 

On the other hand, depending on the system parameters, the relativistic precession of the pericenter can operate on timescales comparable to our integration time of $10^6$ planetary periods and have an effect on the orbital evolution of the planet \citep{2011MNRAS.411..565M,2015ApJ...802...94G}.  The pericentre of Kepler-34 was found to circulate with a period of around $2 \cdot 10^5$ years \citep{2015MNRAS.446.1283C}, while the integration time that corresponds to Kepler-34b, in the context of this work, is around $7.9 \cdot 10^5$ years.  

The aforementioned studies investigated the dynamical evolution of circumbinary planets away from mean motion resonances, by considering mainly secular effects. As we mentioned in section \ref{sec:resu}, one key process leading to three body system instability is the overlap of adjacent mean motion resonances \citep{1980AJ.....85.1122W,2006ApJ...639..423M,2013ApJ...774..129D,2015CeMDA.123..453R,2018AJ....156...95H}. The precise impact of relativity and tidal interactions on circumbinary systems near mean motion resonances is a topic of ongoing research.


\section{Summary and Conclusions} \label{sec:stab}

In this work, we have investigated the problem of dynamical stability of circumbinary orbits subject to Newtonian gravity. 
 We have carried out over 300 million highly accurate numerical simulations of three dimensional and eccentric systems considering a wide range of mass ratios that are relevant to the study of circumbinary planets. Eccentricities of both the stellar binary and the planetary orbits in our study range from 0-0.9. Mass ratios for the inner binary range from equal mass stars down to mass ratios of 0.01. For the outer body, i.e. the circumbinary planet, we explored mass ratios from brown dwarf like bodies down to Mercury mass planets. Our numerical experiments were carried out over a timescale of one million planetary orbital periods. This allows us to capture dynamical effects that require some time to build up (e.g. secular resonances) and affect the planetary orbit.  
 
 We classified the parameter space into three stability regimes separated by critical borders: a zone where investigated orbits were fully stable, a zone where investigated orbits were fully unstable and one zone with mixed behaviour. The mixed zone captures, for instance, the effect of mean motion resonances on the stability of circumbinary planets. We find that the location of the critical stability borders is mainly affected by the eccentricities of the planetary and the binary orbits, the mass ratio of the binary and the mutual inclination between the binary and planetary orbital planes. 
 
 Based on the results of the numerical simulations we derived empirical formulae for the critical semi-major axes of circumbinary planetary systems as a function of the system parameters that identify whether or not a configuration is dynamically stable over a million planetary orbits. We have developed two sets of fitting formulae: one for planetary eccentricities up to 0.8  that is given by equations (\ref{fits}), (\ref{exp1}) and (\ref{exp2})  and one for planetary eccentricities as high as 0.9 that is described by equations (\ref{fits}), (\ref{exp3}) and (\ref{exp4}).
 The former exhibits a better performance in terms of stability classification for most systems, while the latter is meant to cover the entire investigated parameter space.

We have also trained Machine Learning (ML) models on our data set and compared their performance in predicting dynamical stability in circumbinary systems. The empirical formulae and the Machine Learning models were tested against results from numerical simulations of randomly chosen circumbinary systems. 
Both ML and empirical classifiers show similar performance, and constitute a vast improvement over existing techniques, in particular over \cite{1999AJ....117..621H}, in terms of parameter space covered.
%
%

Finally, we tested our empirical stability criteria against known Kepler and TESS systems. The results confirm that many of the circumbinary planets discovered so far orbit near the stability boundary of their respective system.

As our simulations and subsequent results are based on Newtonian gravity and dimensionless mass ratios, they are applicable to any gravitational system compatible with such a model.
Our predictive tools may, thus, be used in various astrophysical contexts, such as planet formation studies, minor planets with moonlets in the solar system as well as the detection and characterization of circumbinary planets in the hunt for habitable worlds.

\begin{acknowledgments}
This research was supported by the Munich Institute for Astro-, Particle and BioPhysics (MIAPbP) which is funded by the Deutsche Forschungsgemeinschaft (DFG, German Research Foundation) under Germany´s Excellence Strategy – EXC-2094 – 390783311. The authors would also like to thank the High Performance Computing Resources team at New York University Abu 
Dhabi, and especially Jorge Naranjo.
This material is based upon work supported by Tamkeen under the NYU Abu
Dhabi Research Institute grant CASS.
\end{acknowledgments}

\bibliography{ref}{}

\begin{thebibliography}{}
\expandafter\ifx\csname natexlab\endcsname\relax\def\natexlab#1{#1}\fi
\providecommand{\url}[1]{\href{#1}{#1}}
\providecommand{\dodoi}[1]{doi:~\href{http://doi.org/#1}{\nolinkurl{#1}}}
\providecommand{\doeprint}[1]{\href{http://ascl.net/#1}{\nolinkurl{http://ascl.net/#1}}}
\providecommand{\doarXiv}[1]{\href{https://arxiv.org/abs/#1}{\nolinkurl{https://arxiv.org/abs/#1}}}

\bibitem[{{Aarseth}(2004)}]{2004RMxAC..21..156A}
{Aarseth}, S.~J. 2004, in Revista Mexicana de Astronomia y Astrofisica
  Conference Series, Vol.~21, Revista Mexicana de Astronomia y Astrofisica
  Conference Series, ed. C.~{Allen} \& C.~{Scarfe}, 156--162

\bibitem[{{Adelbert} {et~al.}(2023){Adelbert}, {Penzlin}, {Sch{\"a}fer},
  {Kley}, {Quarles}, \& {Sfair}}]{2023A&A...680A..29A}
{Adelbert}, S., {Penzlin}, A. B.~T., {Sch{\"a}fer}, C.~M., {et~al.} 2023, \aap,
  680, A29, \dodoi{10.1051/0004-6361/202244329}

\bibitem[{Ali-Dib \& Georgakarakos(2024)}]{mlmodel}
Ali-Dib, M., \& Georgakarakos, N. 2024, {ML model}, 0.1.0,  Zenodo,
  \dodoi{10.5281/zenodo.12628373}

\bibitem[{{Chavez} {et~al.}(2015){Chavez}, {Georgakarakos}, {Prodan},
  {Reyes-Ruiz}, {Aceves}, {Betancourt}, \&
  {Perez-Tijerina}}]{2015MNRAS.446.1283C}
{Chavez}, C.~E., {Georgakarakos}, N., {Prodan}, S., {et~al.} 2015, \mnras, 446,
  1283, \dodoi{10.1093/mnras/stu2142}

\bibitem[{{Chen} {et~al.}(2020){Chen}, {Lubow}, \&
  {Martin}}]{2020MNRAS.494.4645C}
{Chen}, C., {Lubow}, S.~H., \& {Martin}, R.~G. 2020, \mnras, 494, 4645,
  \dodoi{10.1093/mnras/staa1037}

\bibitem[{{Chen} \& {Guestrin}(2016)}]{xgboost}
{Chen}, T., \& {Guestrin}, C. 2016, arXiv e-prints, arXiv:1603.02754,
  \dodoi{10.48550/arXiv.1603.02754}

\bibitem[{{Childs} \& {Martin}(2021)}]{2021MNRAS.507.3461C}
{Childs}, A.~C., \& {Martin}, R.~G. 2021, \mnras, 507, 3461,
  \dodoi{10.1093/mnras/stab2419}

\bibitem[{Chirikov(1979)}]{chirikov1979}
Chirikov, B.~V. 1979, Physics Reports, 52, 263,
  \dodoi{https://doi.org/10.1016/0370-1573(79)90023-1}

\bibitem[{{Correia} {et~al.}(2016){Correia}, {Bou{\'e}}, \&
  {Laskar}}]{2016CeMDA.126..189C}
{Correia}, A. C.~M., {Bou{\'e}}, G., \& {Laskar}, J. 2016, Celestial Mechanics
  and Dynamical Astronomy, 126, 189, \dodoi{10.1007/s10569-016-9709-9}

\bibitem[{{Deck} {et~al.}(2013){Deck}, {Payne}, \&
  {Holman}}]{2013ApJ...774..129D}
{Deck}, K.~M., {Payne}, M., \& {Holman}, M.~J. 2013, \apj, 774, 129,
  \dodoi{10.1088/0004-637X/774/2/129}

\bibitem[{{Doolin} \& {Blundell}(2011)}]{2011MNRAS.418.2656D}
{Doolin}, S., \& {Blundell}, K.~M. 2011, \mnras, 418, 2656,
  \dodoi{10.1111/j.1365-2966.2011.19657.x}

\bibitem[{{Doyle} {et~al.}(2011){Doyle}, {Carter}, {Fabrycky}, {Slawson},
  {Howell}, {Winn}, {Orosz}, {P{\v{r}}sa}, {Welsh}, {Quinn}, {Latham},
  {Torres}, {Buchhave}, {Marcy}, {Fortney}, {Shporer}, {Ford}, {Lissauer},
  {Ragozzine}, {Rucker}, {Batalha}, {Jenkins}, {Borucki}, {Koch}, {Middour},
  {Hall}, {McCauliff}, {Fanelli}, {Quintana}, {Holman}, {Caldwell}, {Still},
  {Stefanik}, {Brown}, {Esquerdo}, {Tang}, {Furesz}, {Geary}, {Berlind},
  {Calkins}, {Short}, {Steffen}, {Sasselov}, {Dunham}, {Cochran}, {Boss},
  {Haas}, {Buzasi}, \& {Fischer}}]{2011Sci...333.1602D}
{Doyle}, L.~R., {Carter}, J.~A., {Fabrycky}, D.~C., {et~al.} 2011, Science,
  333, 1602, \dodoi{10.1126/science.1210923}

\bibitem[{{Dvorak}(1986)}]{1986A&A...167..379D}
{Dvorak}, R. 1986, \aap, 167, 379

\bibitem[{{Dvorak} {et~al.}(1989){Dvorak}, {Froeschle}, \&
  {Froeschle}}]{1989A&A...226..335D}
{Dvorak}, R., {Froeschle}, C., \& {Froeschle}, C. 1989, \aap, 226, 335

\bibitem[{Eggl {et~al.}(2024)Eggl, Georgakarakos, Sakuler, \&
  Pilat-Lohinger}]{exostab2}
Eggl, S., Georgakarakos, N., Sakuler, W., \& Pilat-Lohinger, E. 2024, {Exostab
  2.0}, 0.1.0,  Zenodo, \dodoi{10.5281/zenodo.11243168}

\bibitem[{El-Badry(2024)}]{el2024gaia}
El-Badry, K. 2024, New Astronomy Reviews, 101694

\bibitem[{{Fabrycky} \& {Tremaine}(2007)}]{2007ApJ...669.1298F}
{Fabrycky}, D., \& {Tremaine}, S. 2007, \apj, 669, 1298, \dodoi{10.1086/521702}

\bibitem[{Farago \& Laskar(2010)}]{farago2010high}
Farago, F., \& Laskar, J. 2010, Monthly Notices of the Royal Astronomical
  Society, 401, 1189

\bibitem[{{Georgakarakos}(2008)}]{2008CeMDA.100..151G}
{Georgakarakos}, N. 2008, Celestial Mechanics and Dynamical Astronomy, 100,
  151, \dodoi{10.1007/s10569-007-9109-2}

\bibitem[{{Georgakarakos}(2013)}]{2013NewA...23...41G}
---. 2013, \na, 23, 41, \dodoi{10.1016/j.newast.2013.02.004}

\bibitem[{{Georgakarakos}(2022)}]{2022MNRAS.511.4396G}
---. 2022, \mnras, 511, 4396, \dodoi{10.1093/mnras/stab3332}

\bibitem[{{Georgakarakos} \& {Eggl}(2015)}]{2015ApJ...802...94G}
{Georgakarakos}, N., \& {Eggl}, S. 2015, \apj, 802, 94,
  \dodoi{10.1088/0004-637X/802/2/94}

\bibitem[{{Georgakarakos} {et~al.}(2021){Georgakarakos}, {Eggl}, \&
  {Dobbs-Dixon}}]{2021FrASS...8...44G}
{Georgakarakos}, N., {Eggl}, S., \& {Dobbs-Dixon}, I. 2021, Frontiers in
  Astronomy and Space Sciences, 8, 44, \dodoi{10.3389/fspas.2021.640830}

\bibitem[{{Grishin} {et~al.}(2017){Grishin}, {Perets}, {Zenati}, \&
  {Michaely}}]{2017MNRAS.466..276G}
{Grishin}, E., {Perets}, H.~B., {Zenati}, Y., \& {Michaely}, E. 2017, \mnras,
  466, 276, \dodoi{10.1093/mnras/stw3096}

\bibitem[{{Hadden} \& {Lithwick}(2018)}]{2018AJ....156...95H}
{Hadden}, S., \& {Lithwick}, Y. 2018, \aj, 156, 95,
  \dodoi{10.3847/1538-3881/aad32c}

\bibitem[{Hadjidemetriou(2006)}]{hadjidemetriou2006symmetric}
Hadjidemetriou, J.~D. 2006, Celestial Mechanics and Dynamical Astronomy, 95,
  225

\bibitem[{{Harrington}(1977)}]{1977AJ.....82..753H}
{Harrington}, R.~S. 1977, \aj, 82, 753, \dodoi{10.1086/112121}

\bibitem[{{Holman} \& {Wiegert}(1999)}]{1999AJ....117..621H}
{Holman}, M.~J., \& {Wiegert}, P.~A. 1999, \aj, 117, 621,
  \dodoi{10.1086/300695}

\bibitem[{{Hong} \& {van Putten}(2021)}]{2021NewA...8401516H}
{Hong}, C., \& {van Putten}, M. H.~P.~M. 2021, \na, 84, 101516,
  \dodoi{10.1016/j.newast.2020.101516}

\bibitem[{{Janson} {et~al.}(2012){Janson}, {Hormuth}, {Bergfors}, {Brandner},
  {Hippler}, {Daemgen}, {Kudryavtseva}, {Schmalzl}, {Schnupp}, \&
  {Henning}}]{2012ApJ...754...44J}
{Janson}, M., {Hormuth}, F., {Bergfors}, C., {et~al.} 2012, \apj, 754, 44,
  \dodoi{10.1088/0004-637X/754/1/44}

\bibitem[{{Kenyon} \& {Bromley}(2021)}]{2021AJ....161..211K}
{Kenyon}, S.~J., \& {Bromley}, B.~C. 2021, \aj, 161, 211,
  \dodoi{10.3847/1538-3881/abe858}

\bibitem[{{Kostov} {et~al.}(2013){Kostov}, {McCullough}, {Hinse}, {Tsvetanov},
  {H{\'e}brard}, {D{\'\i}az}, {Deleuil}, \& {Valenti}}]{2013ApJ...770...52K}
{Kostov}, V.~B., {McCullough}, P.~R., {Hinse}, T.~C., {et~al.} 2013, \apj, 770,
  52, \dodoi{10.1088/0004-637X/770/1/52}

\bibitem[{{Kostov} {et~al.}(2014){Kostov}, {McCullough}, {Carter}, {Deleuil},
  {D{\'\i}az}, {Fabrycky}, {H{\'e}brard}, {Hinse}, {Mazeh}, {Orosz},
  {Tsvetanov}, \& {Welsh}}]{2014ApJ...784...14K}
{Kostov}, V.~B., {McCullough}, P.~R., {Carter}, J.~A., {et~al.} 2014, \apj,
  784, 14, \dodoi{10.1088/0004-637X/784/1/14}

\bibitem[{{Kostov} {et~al.}(2016){Kostov}, {Orosz}, {Welsh}, {Doyle},
  {Fabrycky}, {Haghighipour}, {Quarles}, {Short}, {Cochran}, {Endl}, {Ford},
  {Gregorio}, {Hinse}, {Isaacson}, {Jenkins}, {Jensen}, {Kane}, {Kull},
  {Latham}, {Lissauer}, {Marcy}, {Mazeh}, {M{\"u}ller}, {Pepper}, {Quinn},
  {Ragozzine}, {Shporer}, {Steffen}, {Torres}, {Windmiller}, \&
  {Borucki}}]{2016ApJ...827...86K}
{Kostov}, V.~B., {Orosz}, J.~A., {Welsh}, W.~F., {et~al.} 2016, \apj, 827, 86,
  \dodoi{10.3847/0004-637X/827/1/86}

\bibitem[{{Kostov} {et~al.}(2021){Kostov}, {Powell}, {Orosz}, {Welsh},
  {Cochran}, {Collins}, {Endl}, {Hellier}, {Latham}, {MacQueen}, {Pepper},
  {Quarles}, {Sairam}, {Torres}, {Wilson}, {Bergeron}, {Boyce}, {Bieryla},
  {Buchheim}, {Ben Christiansen}, {Ciardi}, {Collins}, {Conti}, {Dixon},
  {Guerra}, {Haghighipour}, {Herman}, {Hintz}, {Howard}, {Jensen}, {Kielkopf},
  {Kruse}, {Law}, {Martin}, {Maxted}, {Montet}, {Murgas}, {Nelson},
  {Olmschenk}, {Otero}, {Quimby}, {Richmond}, {Schwarz}, {Shporer}, {Stassun},
  {Stephens}, {Triaud}, {Ulowetz}, {Walter}, {Wiley}, {Wood}, {Yenawine},
  {Agol}, {Barclay}, {Beatty}, {Boisse}, {Caldwell}, {Christiansen},
  {Col{\'o}n}, {Deleuil}, {Doyle}, {Fausnaugh}, {F{\H{u}}r{\'e}sz}, {Gilbert},
  {H{\'e}brard}, {James}, {Jenkins}, {Kane}, {Kidwell}, {Kopparapu}, {Li},
  {Lissauer}, {Lund}, {Majewski}, {Mazeh}, {Quinn}, {Quintana}, {Ricker},
  {Rodriguez}, {Rowe}, {Santerne}, {Schlieder}, {Seager}, {Standing},
  {Stevens}, {Ting}, {Vanderspek}, \& {Winn}}]{2021AJ....162..234K}
{Kostov}, V.~B., {Powell}, B.~P., {Orosz}, J.~A., {et~al.} 2021, \aj, 162, 234,
  \dodoi{10.3847/1538-3881/ac223a}

\bibitem[{{Lam} \& {Kipping}(2018)}]{2018MNRAS.476.5692L}
{Lam}, C., \& {Kipping}, D. 2018, \mnras, 476, 5692,
  \dodoi{10.1093/mnras/sty022}

\bibitem[{{Liu} {et~al.}(2015){Liu}, {Mu{\~n}oz}, \&
  {Lai}}]{2015MNRAS.447..747L}
{Liu}, B., {Mu{\~n}oz}, D.~J., \& {Lai}, D. 2015, \mnras, 447, 747,
  \dodoi{10.1093/mnras/stu2396}

\bibitem[{{Mardling} \& {Aarseth}(1999)}]{1999ASIC..522..385M}
{Mardling}, R., \& {Aarseth}, S. 1999, in NATO Advanced Study Institute (ASI)
  Series C, Vol. 522, The Dynamics of Small Bodies in the Solar System, A Major
  Key to Solar System Studies, ed. B.~A. {Steves} \& A.~E. {Roy}, 385

\bibitem[{{Mardling}(2008)}]{2008LNP...760...59M}
{Mardling}, R.~A. 2008, in The Cambridge N-Body Lectures, ed. S.~J. {Aarseth},
  C.~A. {Tout}, \& R.~A. {Mardling}, Vol. 760, 59,
  \dodoi{10.1007/978-1-4020-8431-7_3}

\bibitem[{Mardling(2013)}]{mardling2013}
Mardling, R.~A. 2013, Monthly Notices of the Royal Astronomical Society, 435,
  2187, \dodoi{10.1093/mnras/stt1438}

\bibitem[{{Mardling} \& {Aarseth}(2001)}]{2001MNRAS.321..398M}
{Mardling}, R.~A., \& {Aarseth}, S.~J. 2001, \mnras, 321, 398,
  \dodoi{10.1046/j.1365-8711.2001.03974.x}

\bibitem[{{Michtchenko} \& {Malhotra}(2004)}]{2004Icar..168..237M}
{Michtchenko}, T.~A., \& {Malhotra}, R. 2004, \icarus, 168, 237,
  \dodoi{10.1016/j.icarus.2003.12.010}

\bibitem[{{Migaszewski} \& {Go{\'z}dziewski}(2011)}]{2011MNRAS.411..565M}
{Migaszewski}, C., \& {Go{\'z}dziewski}, K. 2011, \mnras, 411, 565,
  \dodoi{10.1111/j.1365-2966.2010.17702.x}

\bibitem[{{Mikkola}(1997)}]{1997CeMDA..67..145M}
{Mikkola}, S. 1997, Celestial Mechanics and Dynamical Astronomy, 67, 145,
  \dodoi{10.1023/A:1008217427749}

\bibitem[{{Mudryk} \& {Wu}(2006)}]{2006ApJ...639..423M}
{Mudryk}, L.~R., \& {Wu}, Y. 2006, \apj, 639, 423, \dodoi{10.1086/499347}

\bibitem[{Naoz {et~al.}(2013)Naoz, Kocsis, Loeb, \& Yunes}]{Naoz_2013}
Naoz, S., Kocsis, B., Loeb, A., \& Yunes, N. 2013, The Astrophysical Journal,
  773, 187, \dodoi{10.1088/0004-637X/773/2/187}

\bibitem[{{Naoz} {et~al.}(2017){Naoz}, {Li}, {Zanardi}, {de El{\'\i}a}, \& {Di
  Sisto}}]{2017AJ....154...18N}
{Naoz}, S., {Li}, G., {Zanardi}, M., {de El{\'\i}a}, G.~C., \& {Di Sisto},
  R.~P. 2017, \aj, 154, 18, \dodoi{10.3847/1538-3881/aa6fb0}

\bibitem[{{Offner} {et~al.}(2022){Offner}, {Moe}, {Kratter}, {Sadavoy},
  {Jensen}, \& {Tobin}}]{2022arXiv220310066O}
{Offner}, S. S.~R., {Moe}, M., {Kratter}, K.~M., {et~al.} 2022, arXiv e-prints,
  arXiv:2203.10066.
\newblock \doarXiv{2203.10066}

\bibitem[{{Orosz} {et~al.}(2012{\natexlab{a}}){Orosz}, {Welsh}, {Carter},
  {Brugamyer}, {Buchhave}, {Cochran}, {Endl}, {Ford}, {MacQueen}, {Short},
  {Torres}, {Windmiller}, {Agol}, {Barclay}, {Caldwell}, {Clarke}, {Doyle},
  {Fabrycky}, {Geary}, {Haghighipour}, {Holman}, {Ibrahim}, {Jenkins},
  {Kinemuchi}, {Li}, {Lissauer}, {Pr{\v{s}}a}, {Ragozzine}, {Shporer}, {Still},
  \& {Wade}}]{2012ApJ...758...87O}
{Orosz}, J.~A., {Welsh}, W.~F., {Carter}, J.~A., {et~al.} 2012{\natexlab{a}},
  \apj, 758, 87, \dodoi{10.1088/0004-637X/758/2/87}

\bibitem[{{Orosz} {et~al.}(2012{\natexlab{b}}){Orosz}, {Welsh}, {Carter},
  {Fabrycky}, {Cochran}, {Endl}, {Ford}, {Haghighipour}, {MacQueen}, {Mazeh},
  {Sanchis-Ojeda}, {Short}, {Torres}, {Agol}, {Buchhave}, {Doyle}, {Isaacson},
  {Lissauer}, {Marcy}, {Shporer}, {Windmiller}, {Barclay}, {Boss}, {Clarke},
  {Fortney}, {Geary}, {Holman}, {Huber}, {Jenkins}, {Kinemuchi}, {Kruse},
  {Ragozzine}, {Sasselov}, {Still}, {Tenenbaum}, {Uddin}, {Winn}, {Koch}, \&
  {Borucki}}]{2012Sci...337.1511O}
---. 2012{\natexlab{b}}, Science, 337, 1511, \dodoi{10.1126/science.1228380}

\bibitem[{{Orosz} {et~al.}(2019){Orosz}, {Welsh}, {Haghighipour}, {Quarles},
  {Short}, {Mills}, {Satyal}, {Torres}, {Agol}, {Fabrycky}, {Jontof-Hutter},
  {Windmiller}, {M{\"u}ller}, {Hinse}, {Cochran}, {Endl}, {Ford}, {Mazeh}, \&
  {Lissauer}}]{2019AJ....157..174O}
{Orosz}, J.~A., {Welsh}, W.~F., {Haghighipour}, N., {et~al.} 2019, \aj, 157,
  174, \dodoi{10.3847/1538-3881/ab0ca0}

\bibitem[{Pilat-Lohinger \& Eggl(2011)}]{pilat2011exostab}
Pilat-Lohinger, E., \& Eggl, S. 2011, Publications of the Astronomy Department
  of the Eotvos Lorand University, 20, 135

\bibitem[{{Pilat-Lohinger} {et~al.}(2003){Pilat-Lohinger}, {Funk}, \&
  {Dvorak}}]{2003A&A...400.1085P}
{Pilat-Lohinger}, E., {Funk}, B., \& {Dvorak}, R. 2003, \aap, 400, 1085,
  \dodoi{10.1051/0004-6361:20021811}

\bibitem[{{Popova} \& {Shevchenko}(2016)}]{2016AstL...42..474P}
{Popova}, E.~A., \& {Shevchenko}, I.~I. 2016, Astronomy Letters, 42, 474,
  \dodoi{10.1134/S1063773716060050}

\bibitem[{{Prodan} \& {Murray}(2012)}]{2012ApJ...747....4P}
{Prodan}, S., \& {Murray}, N. 2012, \apj, 747, 4,
  \dodoi{10.1088/0004-637X/747/1/4}

\bibitem[{{Quarles} {et~al.}(2020){Quarles}, {Li}, {Kostov}, \&
  {Haghighipour}}]{2020AJ....159...80Q}
{Quarles}, B., {Li}, G., {Kostov}, V., \& {Haghighipour}, N. 2020, \aj, 159,
  80, \dodoi{10.3847/1538-3881/ab64fa}

\bibitem[{{Quarles} {et~al.}(2018){Quarles}, {Satyal}, {Kostov}, {Kaib}, \&
  {Haghighipour}}]{2018ApJ...856..150Q}
{Quarles}, B., {Satyal}, S., {Kostov}, V., {Kaib}, N., \& {Haghighipour}, N.
  2018, \apj, 856, 150, \dodoi{10.3847/1538-4357/aab264}

\bibitem[{{Raghavan} {et~al.}(2010){Raghavan}, {McAlister}, {Henry}, {Latham},
  {Marcy}, {Mason}, {Gies}, {White}, \& {ten Brummelaar}}]{2010ApJS..190....1R}
{Raghavan}, D., {McAlister}, H.~A., {Henry}, T.~J., {et~al.} 2010, \apjs, 190,
  1, \dodoi{10.1088/0067-0049/190/1/1}

\bibitem[{{Ramos} {et~al.}(2015){Ramos}, {Correa-Otto}, \&
  {Beaug{\'e}}}]{2015CeMDA.123..453R}
{Ramos}, X.~S., {Correa-Otto}, J.~A., \& {Beaug{\'e}}, C. 2015, Celestial
  Mechanics and Dynamical Astronomy, 123, 453,
  \dodoi{10.1007/s10569-015-9646-z}

\bibitem[{{Schwamb} {et~al.}(2013){Schwamb}, {Orosz}, {Carter}, {Welsh},
  {Fischer}, {Torres}, {Howard}, {Crepp}, {Keel}, {Lintott}, {Kaib}, {Terrell},
  {Gagliano}, {Jek}, {Parrish}, {Smith}, {Lynn}, {Simpson}, {Giguere}, \&
  {Schawinski}}]{2013ApJ...768..127S}
{Schwamb}, M.~E., {Orosz}, J.~A., {Carter}, J.~A., {et~al.} 2013, \apj, 768,
  127, \dodoi{10.1088/0004-637X/768/2/127}

\bibitem[{{Shevchenko}(2015)}]{2015ApJ...799....8S}
{Shevchenko}, I.~I. 2015, \apj, 799, 8, \dodoi{10.1088/0004-637X/799/1/8}

\bibitem[{{Socia} {et~al.}(2020){Socia}, {Welsh}, {Orosz}, {Cochran}, {Endl},
  {Quarles}, {Short}, {Torres}, {Windmiller}, \&
  {Yenawine}}]{2020AJ....159...94S}
{Socia}, Q.~J., {Welsh}, W.~F., {Orosz}, J.~A., {et~al.} 2020, \aj, 159, 94,
  \dodoi{10.3847/1538-3881/ab665b}

\bibitem[{{Standing} {et~al.}(2023){Standing}, {Sairam}, {Martin}, {Triaud},
  {Correia}, {Coleman}, {Baycroft}, {Kunovac}, {Boisse}, {Cameron},
  {Dransfield}, {Faria}, {Gillon}, {Hara}, {Hellier}, {Howard}, {Lane},
  {Mardling}, {Maxted}, {Miller}, {Nelson}, {Orosz}, {Pepe}, {Santerne},
  {Sebastian}, {Udry}, \& {Welsh}}]{2023NatAs...7..702S}
{Standing}, M.~R., {Sairam}, L., {Martin}, D.~V., {et~al.} 2023, Nature
  Astronomy, 7, 702, \dodoi{10.1038/s41550-023-01948-4}

\bibitem[{{Sundman}(1913)}]{sund}
{Sundman}, K. 1913, Acta Math., 36, 105, \dodoi{10.1007/BF02422379}

\bibitem[{{Szebehely}(1984)}]{1984CeMec..34...49S}
{Szebehely}, V. 1984, Celestial Mechanics, 34, 49, \dodoi{10.1007/BF01235791}

\bibitem[{{Tory} {et~al.}(2022){Tory}, {Grishin}, \&
  {Mandel}}]{2022PASA...39...62T}
{Tory}, M., {Grishin}, E., \& {Mandel}, I. 2022, \pasa, 39, e062,
  \dodoi{10.1017/pasa.2022.57}

\bibitem[{{Verrier} \& {Evans}(2009)}]{2009MNRAS.394.1721V}
{Verrier}, P.~E., \& {Evans}, N.~W. 2009, \mnras, 394, 1721,
  \dodoi{10.1111/j.1365-2966.2009.14446.x}

\bibitem[{{Vynatheya} {et~al.}(2022){Vynatheya}, {Hamers}, {Mardling}, \&
  {Bellinger}}]{2022MNRAS.516.4146V}
{Vynatheya}, P., {Hamers}, A.~S., {Mardling}, R.~A., \& {Bellinger}, E.~P.
  2022, \mnras, 516, 4146, \dodoi{10.1093/mnras/stac2540}

\bibitem[{{Welsh} {et~al.}(2012){Welsh}, {Orosz}, {Carter}, {Fabrycky}, {Ford},
  {Lissauer}, {Pr{\v{s}}a}, {Quinn}, {Ragozzine}, {Short}, {Torres}, {Winn},
  {Doyle}, {Barclay}, {Batalha}, {Bloemen}, {Brugamyer}, {Buchhave},
  {Caldwell}, {Caldwell}, {Christiansen}, {Ciardi}, {Cochran}, {Endl},
  {Fortney}, {Gautier}, {Gilliland}, {Haas}, {Hall}, {Holman}, {Howard},
  {Howell}, {Isaacson}, {Jenkins}, {Klaus}, {Latham}, {Li}, {Marcy}, {Mazeh},
  {Quintana}, {Robertson}, {Shporer}, {Steffen}, {Windmiller}, {Koch}, \&
  {Borucki}}]{2012Natur.481..475W}
{Welsh}, W.~F., {Orosz}, J.~A., {Carter}, J.~A., {et~al.} 2012, \nat, 481, 475,
  \dodoi{10.1038/nature10768}

\bibitem[{{Welsh} {et~al.}(2015){Welsh}, {Orosz}, {Short}, {Cochran}, {Endl},
  {Brugamyer}, {Haghighipour}, {Buchhave}, {Doyle}, {Fabrycky}, {Hinse},
  {Kane}, {Kostov}, {Mazeh}, {Mills}, {M{\"u}ller}, {Quarles}, {Quinn},
  {Ragozzine}, {Shporer}, {Steffen}, {Tal-Or}, {Torres}, {Windmiller}, \&
  {Borucki}}]{2015ApJ...809...26W}
{Welsh}, W.~F., {Orosz}, J.~A., {Short}, D.~R., {et~al.} 2015, \apj, 809, 26,
  \dodoi{10.1088/0004-637X/809/1/26}

\bibitem[{{Wisdom}(1980)}]{1980AJ.....85.1122W}
{Wisdom}, J. 1980, \aj, 85, 1122, \dodoi{10.1086/112778}

\end{thebibliography}
\bibliographystyle{aasjournal}

\end{document}